\documentclass{article}

\usepackage{epsfig}
\usepackage{graphics}
\usepackage{graphicx}
\usepackage[centertags]{amsmath}
\usepackage{amsfonts}
\usepackage{amsthm}
\usepackage{amssymb}
\usepackage{url}
\usepackage{subfigure}
\usepackage{hyperref}

\usepackage{color}

\newenvironment{myitemize}{
\begin{itemize}
  \setlength{\itemsep}{1pt}
  \setlength{\parskip}{0pt}
  \setlength{\parsep}{0pt}}{\end{itemize}
}

\begin{document}

\title{Wolfram's Classification and Computation in Cellular Automata Classes III and IV}

\author{Genaro J. Mart{\'i}nez$^{1}$, Juan C. Seck-Tuoh-Mora$^2$, and Hector Zenil$^3$}
\date{}
%begin \date{January 23, 2012}

\maketitle

\begin{centering}
$^1$ Unconventional Computing Center, Bristol Institute of Technology, University of the West of England, Bristol, UK.\\
Departamento de Ciencias e Ingenier{\'i}a de la Computaci{\'o}n, Escuela Superior de C\'omputo, Instituto Polit\'ecnico Nacional, M\'exico.
\url{genaro.martinez@uwe.ac.uk}\\
$^2$ Centro de Investigaci\'on Avanzada en Ingenier\'ia Industrial\\Universidad Aut\'onoma del Estado de Hidalgo, M\'exico.\\
\url{jseck@uaeh.edu.mx} \\
$^3$ Behavioural and Evolutionary Theory Lab\\Department of Computer Science, University of Sheffield, UK.\\
\url{h.zenil@sheffield.ac.uk} \\
\end{centering}

\begin{abstract}
We conduct a brief survey on Wolfram's classification, in particular related to the computing capabilities of Cellular Automata (CA) in Wolfram's classes III and IV. We formulate and shed light on the question of whether Class III systems are capable of Turing universality or may turn out to be ``too hot'' in practice to be controlled and programmed. We show that systems in Class III are indeed capable of computation and that there is no reason to believe that they are unable, in principle, to reach Turing-completness.\\

\end{abstract}

\section{Wolfram's classification of Cellular Automata}

A comment in Wolfram's {\it A New Kind of Science} gestures toward the first difficult problem we will tackle (ANKOS) (page 235): {\it trying to predict detailed properties of a particular cellular automaton, it was often enough just to know what class the cellular automaton was in}. The second problem we will take on concerns the possible relation between complexity of Cellular Automata and Turing universal computation, also highlighted by Wolfram in his ANKOS (page 691-- on Class 4 behaviour and Universality): {\it I strongly suspect that it is true in general that any cellular automaton which shows overall class 4 behaviour will turn out---like Rule 110---to be universal}. \index{Rule 110}
\index{cellular automata}\index{Wolfram's classification}
The classification and identification of cellular automata (CA) has become a central focus of research in the field. In \cite{kn:Wolf84}, Stephen Wolfram presented his now well-known {\it classes}. Wolfram's analysis included a thorough study of one-dimensional (1D) CA, order $(k=2,r=2)$ (where $k \in \mathcal{Z^+}$ is the cardinality of the finite alphabet and $r \in \mathcal{Z^+}$ the number of neighbours), and also found the same classes of behaviour in other CA rule spaces. This allowed Wolfram to generalise his classification to all sorts of systems in \cite{kn:Wolf02}. 

An Elementary Cellular Automaton (ECA) is a finite automaton defined in a 1D array. The automaton assumes two states, and updates its state in discrete time according to its own state and the state of its two closest neighbours, all cells updating their states synchronously. \\

Wolfram's classes can be characterised as follows:

\begin{myitemize}
\item {Class I.} CA evolving to a homogeneous state
\item {Class II.} CA evolving periodically
\item {Class III.} CA evolving chaotically
\item {Class IV.} Includes all previous cases, known as a class of {\it complex rules}
\end{myitemize}

Otherwise explained, in the case of a given CA,:

\begin{myitemize}
\item If the evolution is dominated by a unique state of its alphabet for any random initial condition, then it belongs to {\it Class I}.
\item If the evolution is dominated by blocks of cells which are periodically repeated for any random initial condition, then it belongs to {\it Class II}.
\item If for a long time and for any random initial condition, the evolution is dominated by sets of cells without any defined pattern, then it belongs to {\it Class III}.
\item If the evolution is dominated by non-trivial structures emerging and travelling along the evolution space where uniform, periodic, or chaotic regions can coexist with these structures, then it belongs to {\it Class IV}. This class is frequently tagged: {\it complex behaviour}, {\it complexity dynamics}, or simply {\it complex}.
\end{myitemize} 

\begin{figure}%[th]
\begin{center}
\subfigure[]{\scalebox{0.2}{\includegraphics{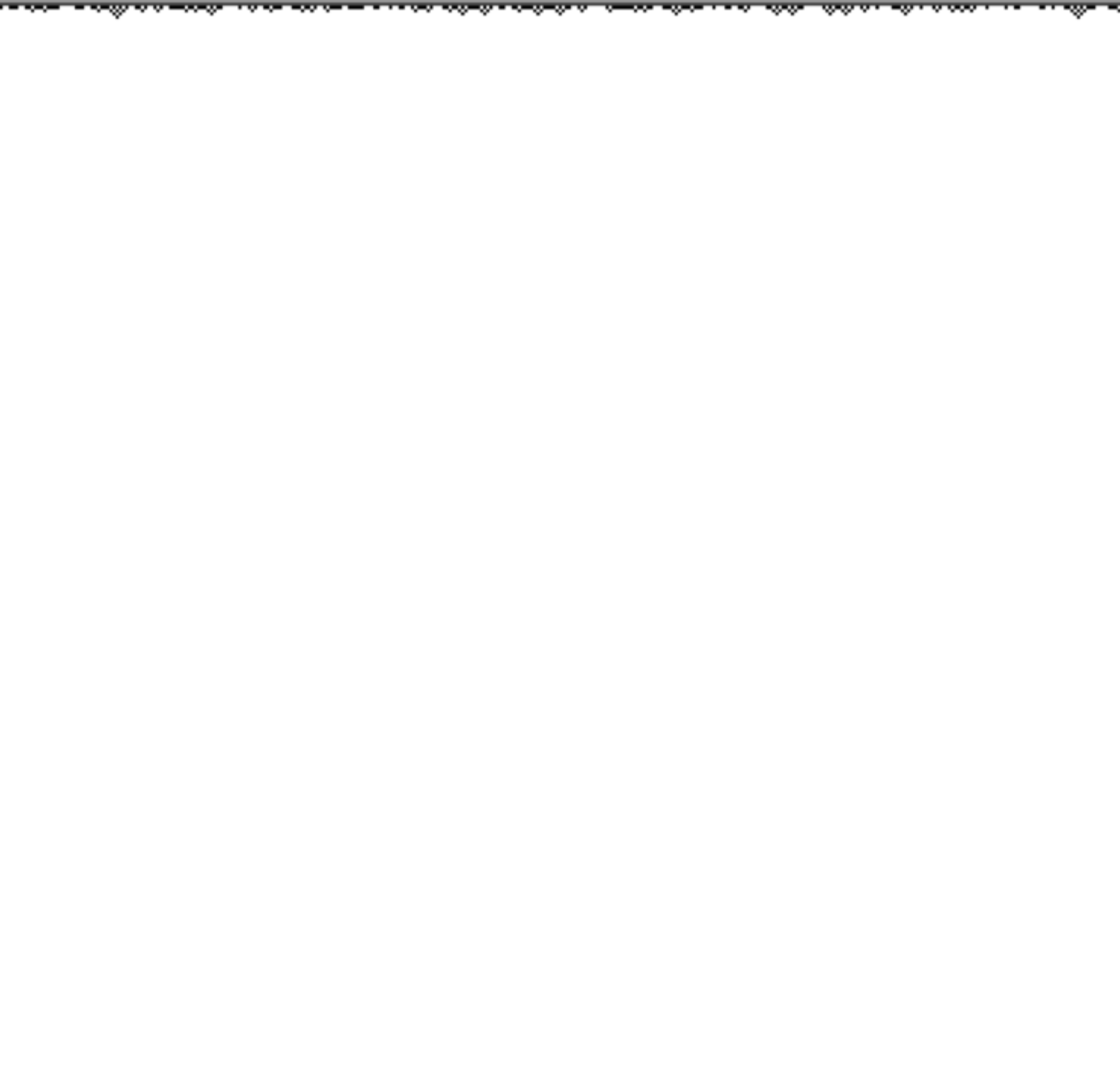}}} %\hspace{0.1cm}
\subfigure[]{\scalebox{0.2}{\includegraphics{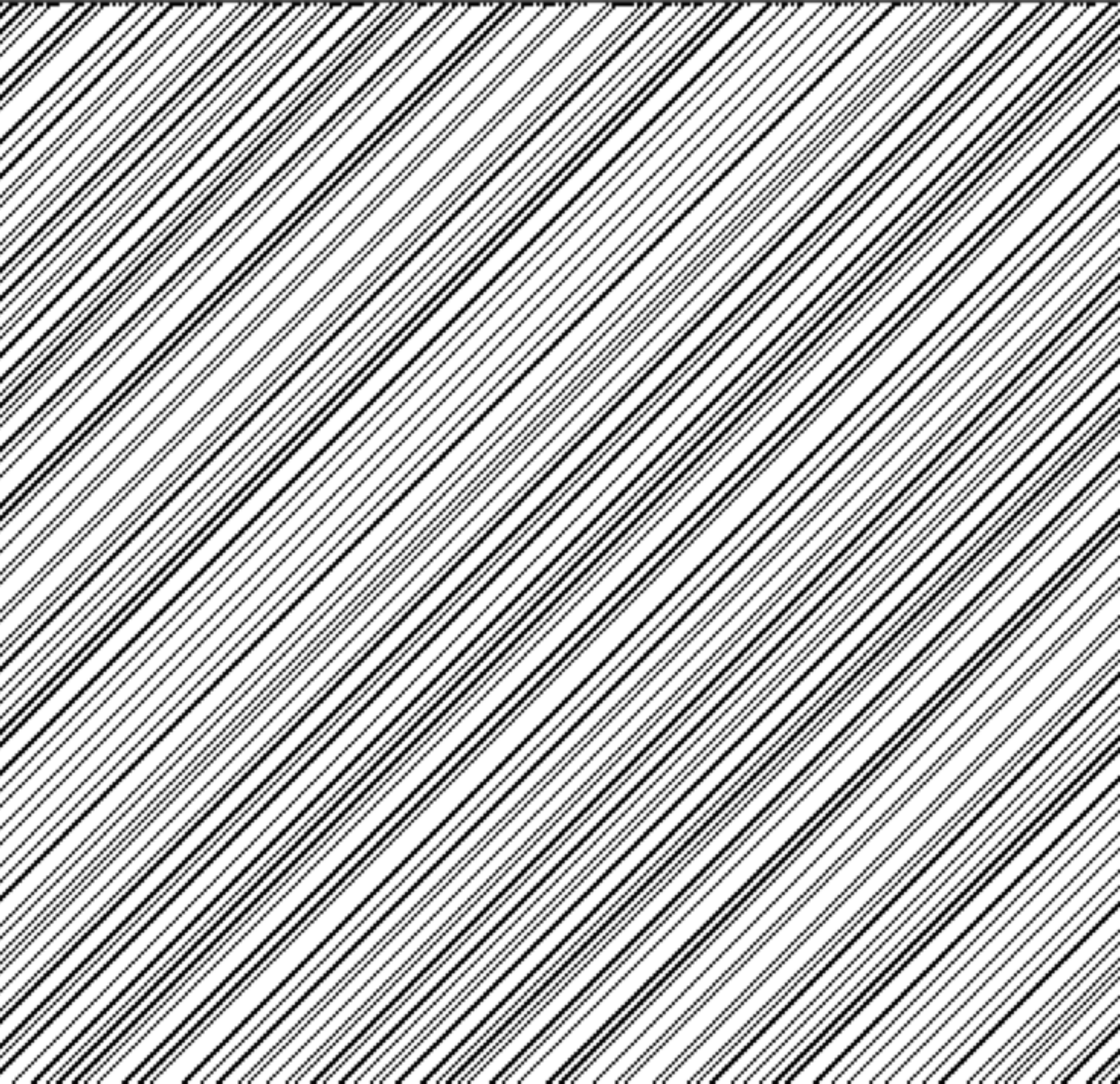}}} %\hspace{0.1cm}
\subfigure[]{\scalebox{0.2}{\includegraphics{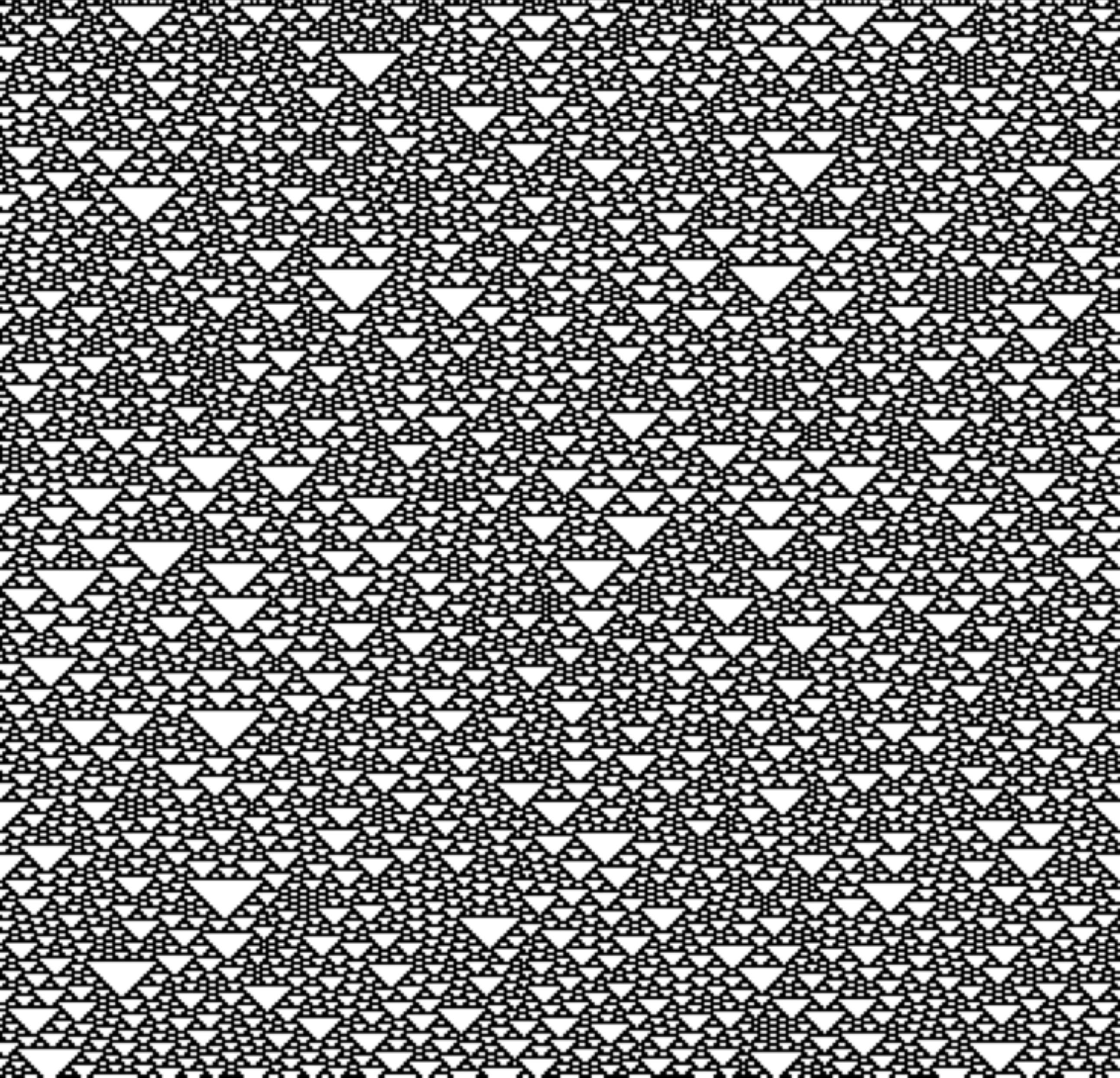}}} %\hspace{0.1cm}
\subfigure[]{\scalebox{0.2}{\includegraphics{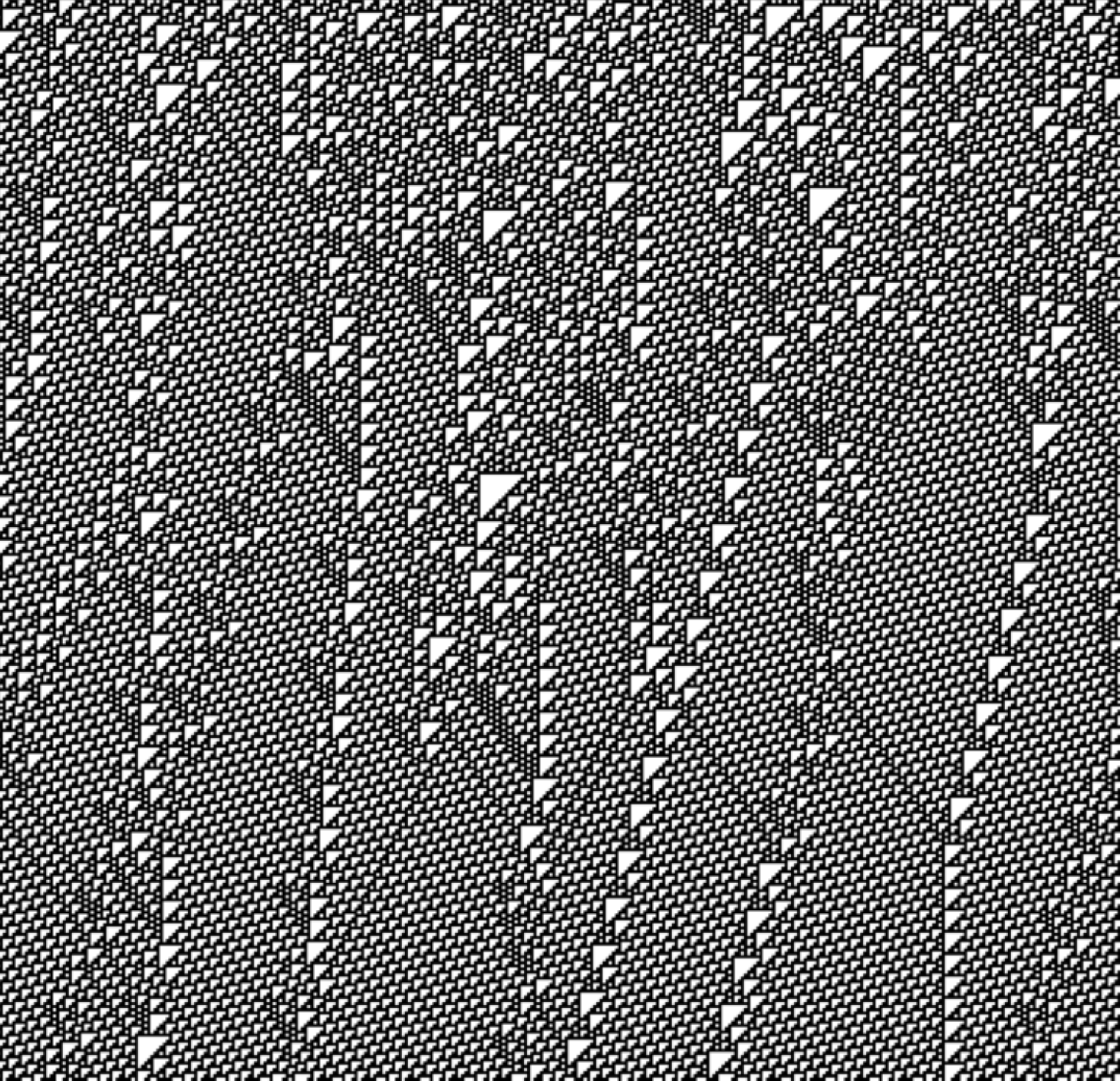}}} \end{center}
\caption{Wolfram's classes represented by ECA rules: (a) Class I - ECA Rule 32, (b) Class II - ECA Rule 10, (c) Class III - ECA Rule 126, (d) Class IV - ECA Rule 110. We have the same initial condition in all these cases, with a density of 50\% for state 0 (white dots) and state 1 (black dots). The evolution space begins with a ring of 358 cells for 344 generations.}
\label{WolframClasses}
\end{figure}

Fig.~\ref{WolframClasses} illustrates Wolfram's classes, focusing on a specific ECA evolution rule (following  Wolfram's notation for ECA \cite{kn:Wolf83}). All evolutions begin with the same random initial condition. Thus, Fig.~\ref{WolframClasses}a displays ECA Rule 32 converging quickly to a homogeneous state, Class I. Figure~\ref{WolframClasses}b displays blocks of cells in state one which evolve periodically showing a leftward shift, Class II. Figure~\ref{WolframClasses}c displays a typical chaotic evolution, where no pattern can be recognised or any limit point identified, Class III. Finally, Fig.~\ref{WolframClasses}d displays the so called complex class or Class IV. Here we see non-trivial patterns emerging in the evolution space. Such patterns possess a defined form and travel along the evolution space. They interact (collide), giving rise to interesting reactions such as annihilations, fusions, solitons and reflections, or they produce new structures. These patterns are referred to as {\it gliders} in the CA literature (`glider' is a widely accepted concept popularised by John Conway through his well-known additive binary 2D CA, the {\it Game of Life} (GoL) \cite{kn:Gard70}). In Class IV CA we see regions with periodic evolutions and chaos, and most frequently in complex rules the background is dominated by stable states, such as in GoL. In such cases---and this is particularly true of the complex ECA Rule 110--the CA can evolve with a periodic background (called ether) where these gliders emerge and live. Gliders in GoL and other CAs such as the 2D Brian's Brain CA \cite{kn:TM87} caught the attention of Christopher Langton, spurring the development of the field of {\it Artificial Life} (AL) \cite{kn:Lang84, kn:Lang86}. 

Since the publication of the paper ``Universality and complexity in cellular automata'' in 1984 \cite{kn:Wolf84},  qualitative classifications of CA (an other systems) have been a much studied and disputed subject. Wolfram advanced several ECA rules as representatives for each of his classes and despite an early comment suggesting that (page 31): {\it $k=2, r=1$ cellular automata are too simple to support universal computation}, in his book ``Cellular Automata and Complexity'' \cite{kn:Wolf94} ECA Rule 110 was granted its own appendix (Table 15, Structures in Rule 110, pages 575--577). It contains specimens of evolutions, including a list of thirteen gliders compiled by Doug Lind, and also presents the conjecture that the rule could be universal. Wolfram writes: {\it One may speculate that the behaviour of Rule 110 is sophisticated enough to support universal computation}. 

An interesting paper written by Karel Culik II and Sheng Yu titled ``Undecidability of CA Classification Schemes'' \cite{kn:CY88, kn:Sut89} discussed the properties of such classes, concluding that: {\it it is undecidable to which class a given cellular automaton belongs} (page 177). Indeed, in 1984 Wolfram \cite{kn:Wolf84} commented (page 1): {\it The fourth class is probably capable of universal computation, so that properties of its infinite time behaviour are undecidable}. Actually, we can see that no effective algorithm exists that is capable of deciding whether a CA is complex or universal, and so far only a few discovered (as opposed to constructed) cellular automata have been proven to be capable of universal computation (notably Wolfram's Rule 110 and Conway's Game of Life). However some techniques offer suitable approximations for finding certain sets of complex, though perhaps not necessarily universal rules (under Wolfram's PCE they would be, c.f. Section \ref{finalremarks}).
\index{Rule 30}
In \cite{israeli}, Israeli and Goldenfeld devised a coarse-grained technique to find predictable properties of elementary CA and other systems. While they were able to reduce elementary CAs in all Wolfram's classes, they were unable to do so for some in Class III (rules 30, 45, 106 and their symmetries) and more surprisingly in Class II (rule 154 and its symmetries). Their technique showed to be able to find properties of CA at some coarse-grained level of description without accounting for small-scale details. They show that by using this technique one can take a Class III system to a Class I in order to predict some properties of the original system by a reduction of its apparent complexity, pointing out that irreducibility may not be the same as complexity (or universality) given that some irreducible rules can be coarse-grained (at least one example of an irreducible rule (110) is known for certain because its ability of Turing universality). This seems in agreement with the fact that systems in Class IV seem to show more persistent structures than systems in Class III.
\index{Gutowitz, H.}
In ``Local structure theory for cellular automata'' \cite{kn:GVK87} Howard Gutowitz developed a statistical analysis. An interesting schematic diagram conceptualising the umbral of classes of CA was offered by Wentian Li and Norman Packard in ``The Structure of the Elementary Cellular Automata Rule Space'' \cite{kn:LP90}. Pattern recognition and classification has been examined in ``Toward the classification of the patterns generated by one-dimensional cellular automata'' \cite{kn:AN86} by Yoji Aizawa\index{Aizawa, Y.} and \index{Ikuko, N.} Ikuko Nishikawa. An extended analysis by Andrew Adamatzky\index{Adamatzky, A.} under the heading ``Identification of Cellular Automata'' in \cite{kn:Ada94} considered the problem of how, given a sequence of configurations of an unknown cellular automaton, one may reconstruct its evolution rules. A recent special issue dedicated to this problem focuses on some theoretical and practical results.\footnote{Special issue ``Identification of Cellular Automata'', \textit{Journal of Cellular Automata} {\bf 2(1)}, 1--102, 2007. \url{http://www.oldcitypublishing.com/JCA/JCAcontents/JCAv2n1contents.html}} Klaus Sutner\index{Sutner, K} has discussed this classification and also the principle of computational equivalence in ``Classification of Cellular Automata'' \cite{kn:Sut09}, with an emphasis on Class IV or computable CA. An interesting approach involving an additive 2D CA was described in \index{Eppstein, D.}David Eppstein's classification scheme \cite{kn:Epp99}\footnote{For a discussion see Tim Tyler's CA FAQ at \url{http://cafaq.com/classify/index.php}, and more recently, a compression-based technique inspired by algorithmic information theory has been advanced\cite{kn:Zen10} that offers a powerful method for identifying complex CA and other complex systems }. 

We will discuss some practical and theoretical topics that distinguish such classes and explore the computing properties of CA rules, in particular in classes III and IV. Among the topics we want to explore is the feasibility of using extended analog computers (EAC) \cite{kn:Mills08} for CA construction, in order to obtain unconventional computing models \cite{kn:Ada02, kn:Ada01}. In this classification, Class IV is of particular interest because the rules of the class present non-trivial behaviour, with a rich diversity of patterns emerging, and non-trivial interactions between gliders, plus mobile localizations, particles, or fragments of waves. This feature was useful in implementing a register machine in GoL \cite{kn:BCG82} to determine its universality. First we survey some of the approximations that allow the identification of complex properties of CA and other systems.

%%%%%%%%%%%%%%%%%%%%%
\subsection{Mean field approximation}
\index{Mean field approximation}
The Mean field theory is a well-known technique for discovering the statistical properties of CA without analysing the evolution space of individual rules. It has been used extensively by Gutowitz in \cite{kn:Guto89}. The method assumes that states in $\Sigma$ are independent and do not correlate with each other in the local function $\varphi$. Thus we can study probabilities of states in a neighbourhood in terms of the probability of a single state (the state in which the neighbourhood evolves), and the probability of a neighbourhood would be the product of the probabilities of each cell in it.

Harold V. McIntosh in \cite{kn:Mc90} presents an explanation of Wolfram's classes using a mixture of probability theory and de Bruijn diagrams\footnote{The de Bruijn diagrams have been culled from Masakazu Nasu's 1978 work on tessellation automata \cite{kn:Nasu78}. Wolfram himself has explored some of this in  \cite{kn:Wolf84a}, later thoroughly analysed by McIntosh \cite{kn:Mc91, kn:Mc09}, Sutner \cite{kn:Sut91}, Burton Voorhes \cite{kn:Voor96, kn:Voor06}, and, particularly, exploited to calculate reversible 1D CA using de Bruijn diagrams derived from the Welch diagrams by Seck-Tuoh-Mora in \cite{kn:SCM05, kn:SMM06}}, resulting in a classification based on the mean field theory curve:

\begin{myitemize}
\item Class I: monotonic, entirely on one side of diagonal;
\item Class II: horizontal tangency, never reaches diagonal;
\item Class IV: horizontal plus diagonal tangency, no crossing; 
\item Class III: no tangencies, curve crosses diagonal.
\end{myitemize}

For the one-dimensional case, all neighbourhoods are considered, as follows:

\begin{equation}
p_{t+1}=\sum_{j=0}^{k^{2r+1}-1}\varphi_{j}(X)p_{t}^{v}(1-p_{t})^{n-v}
\label{MFp1D}
\end{equation}

\noindent such that $j$ indexes every neighbourhood, $X$ are cells $x_{i-r}, \ldots, x_{i}, \ldots, x_{i+r}$, $n$ is the number of cells in every neighbourhood, $v$ indicates how often state `1' occurs in $X$, $n-v$ shows how often state `0' occurs in the neighbourhood $X$, $p_{t}$ is the probability of a cell being in state `1', and $q_{t}$ is the probability of a cell being in state `0'; i.e., $q=1-p$. For Mean field theory in other lattices and dimensions, please consult\cite{kn:Guto87a, kn:Guto89a}.

\index{Conway's Game of Life}
%%%%%%%%%%%%%%%%%%%%%
\subsection{Basins of attraction approximation}\index{Lesser, M.}\index{Wuensche, A.}
Andrew Wuensche, together with Mike Lesser, published a landmark book entitled ``The Global Dynamics of Cellular Automata'' in 1992 \cite{kn:WL92} which contained a very extended analysis of attractors in ECA. Wolfram himself had explored part of these cycles in ``Random Sequence Generation by Cellular Automata'' \cite{kn:Wolf86}, as had McIntosh in ``One Dimensional Cellular Automata'' \cite{kn:Mc09}. Notably, Stuart Kauffman in his book ``The Origins of Order: Self-Organization and Selection in Evolution'' \cite{kn:Kau93} applies basins of attraction to sample random Boolean networks (RBN) in order to illustrate his idea that RBN constitute a model of the gene regulatory network, and that cell types are attractors. The best description of such an analysis is to be found in \cite{kn:Wue98}. 
\index{self-organisation}
A basin (of attraction) field of a finite CA is the set of basins of attraction into which all possible states and trajectories will be organized by the local function $\varphi$. The topology of a single basin of attraction may be represented by a diagram, the {\it state transition graph}. Thus the set of graphs composing the field specifies the global behaviour of the system \cite{kn:WL92}. 
\index{basins of attraction}
Generally a basin can also recognize CA with chaotic or complex behaviour using prior results on attractors \cite{kn:WL92}. Thus, Wuensche says that Wolfram's classes can be represented as a {\it basin classification} \cite{kn:WL92}, as follows:

\begin{myitemize}
\item Class I: very short transients, mainly point attractors (but possibly also periodic attractors), very high in-degree, very high leaf density (very ordered dynamics);
\item Class II: very short transients, mainly short periodic attractors (but also point attractors), high in-degree, very high leaf density;
\item Class IV: moderate transients, moderate-length periodic attractors, moderate in-degree, very moderate leaf density (possibly complex dynamics);
\item Class III: very long transients, very long periodic attractors, low in-degree, low leaf density (chaotic dynamics).
\end{myitemize}

\subsection{Compressibility approximation}
\label{compressibility} 
\index{Compressibility approximation}
\index{Kolmogorov complexity}

A compression-based classification of CA (and other systems) was proposed in \cite{kn:Zen10}, based on the concept of algorithmic (Kolmogorov) complexity. Unlike the Mean field theory, this technique analyses the asymptotic statistical properties of CA by looking at full space-time evolution of individual rules up to an arbitrary number of steps. The method produces the following variation of Wolfram's classification \cite{kn:ZenAISB}. 

\begin{myitemize}
\item Class I: highly compressible evolutions for any number of steps;
\item Class II: highly compressible evolutions for any number of steps;
\item Class III: the lengths of compressed evolutions asymptotically converge to the uncompressed evolution lengths;
\item Class IV: the lengths of compressed evolutions asymptotically converge to the uncompressed evolution lengths.
\end{myitemize}

The four classes seem to give way to only two (Classes I and II and Classes III and IV are not distinguishable in this first approach). But it is shown how algorithmic information theory helps to separate them again, using the concept of asymptotic behaviour advanced in \cite{kn:Zen10,kn:Zen12}.

\begin{figure}%[th]
\begin{center}
\subfigure[]{\scalebox{0.265}{\includegraphics{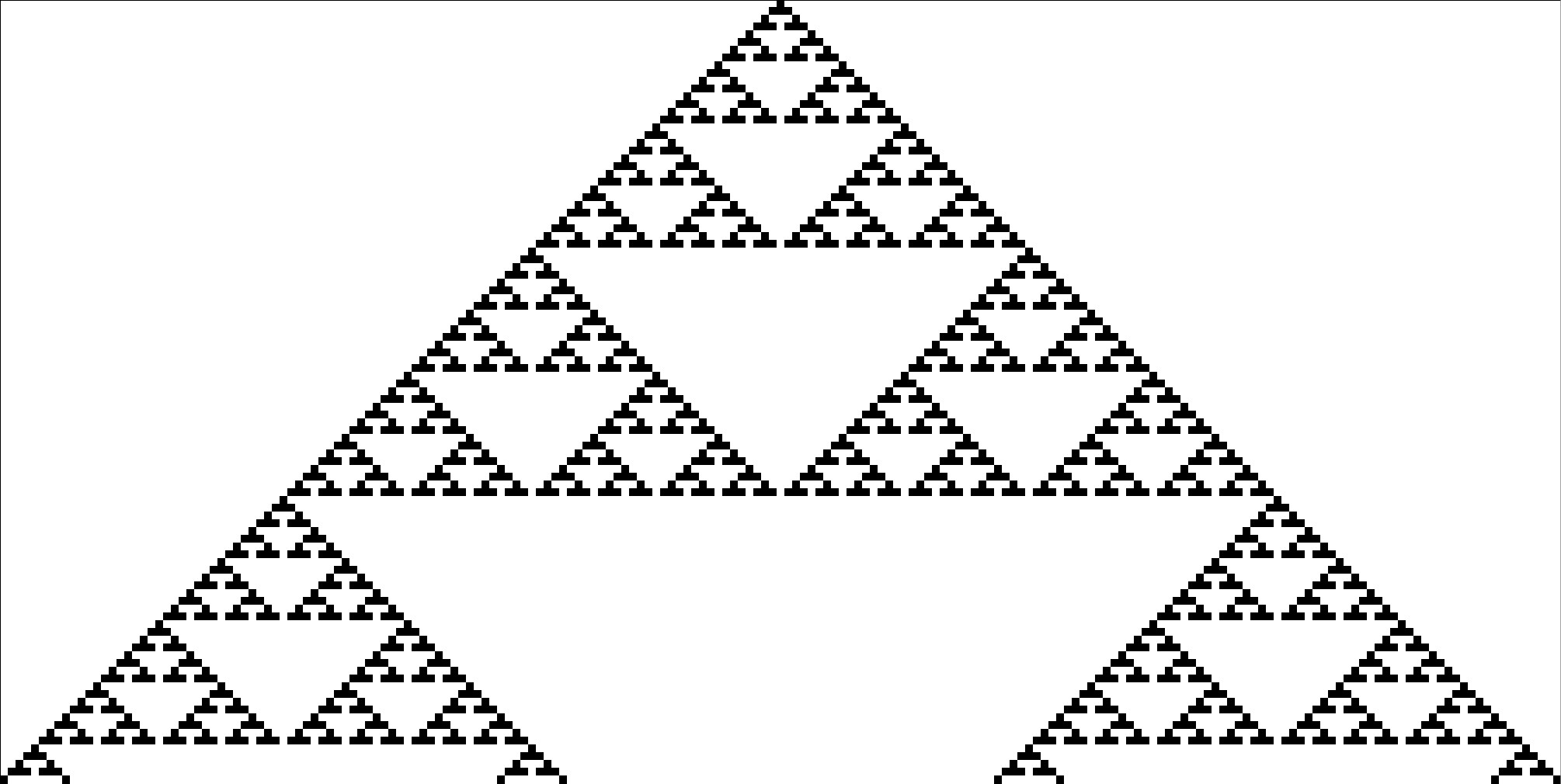}}} 
\subfigure[]{\scalebox{0.265}{\includegraphics{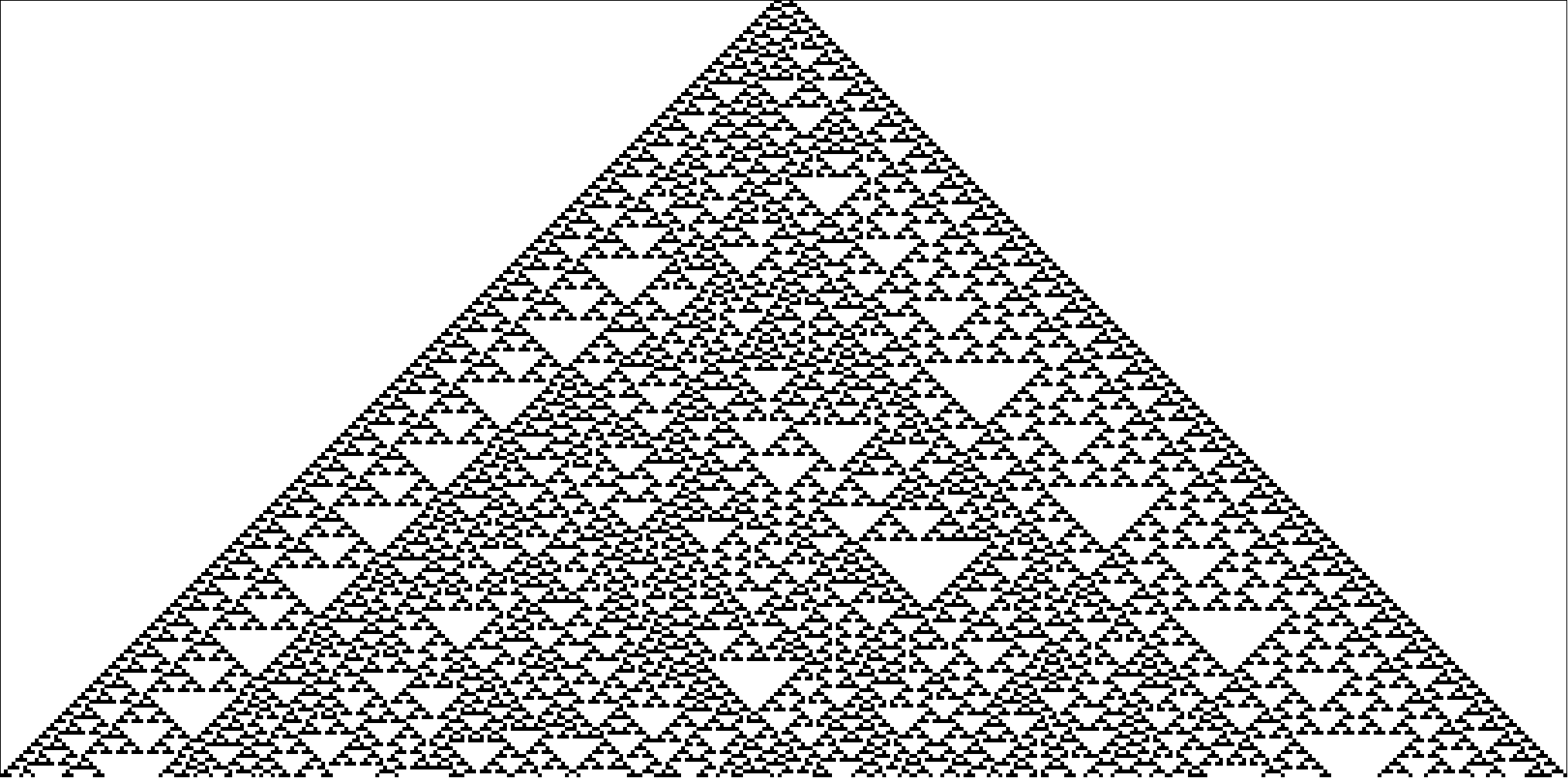}}}
\subfigure[]{\scalebox{0.44}{\includegraphics{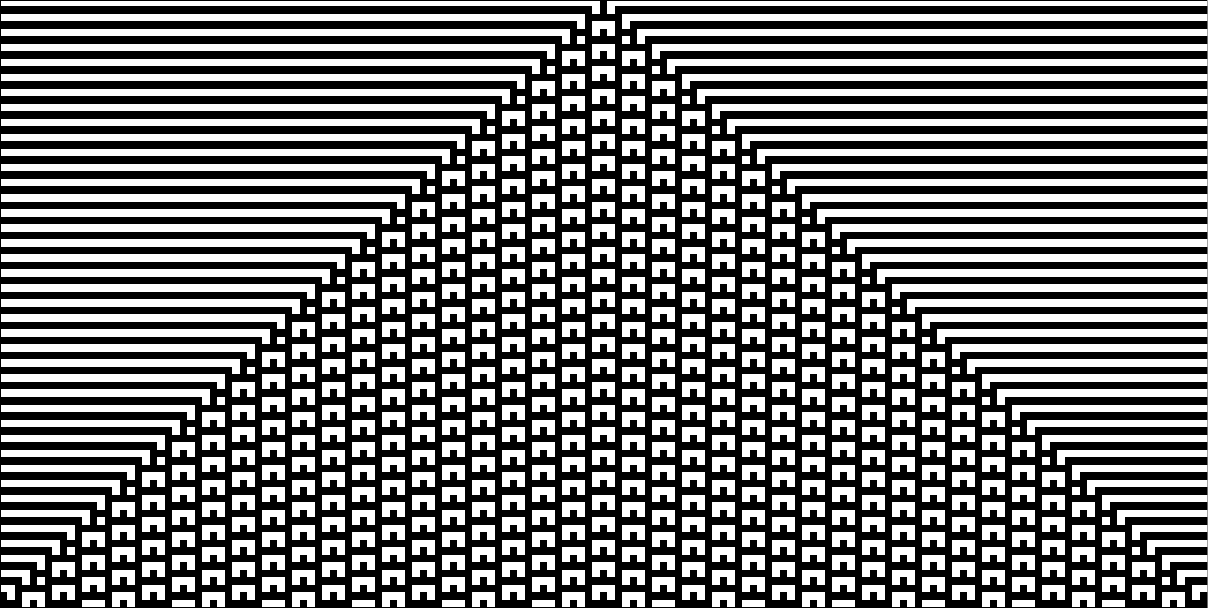}}} 
\subfigure[]{\scalebox{0.44}{\includegraphics{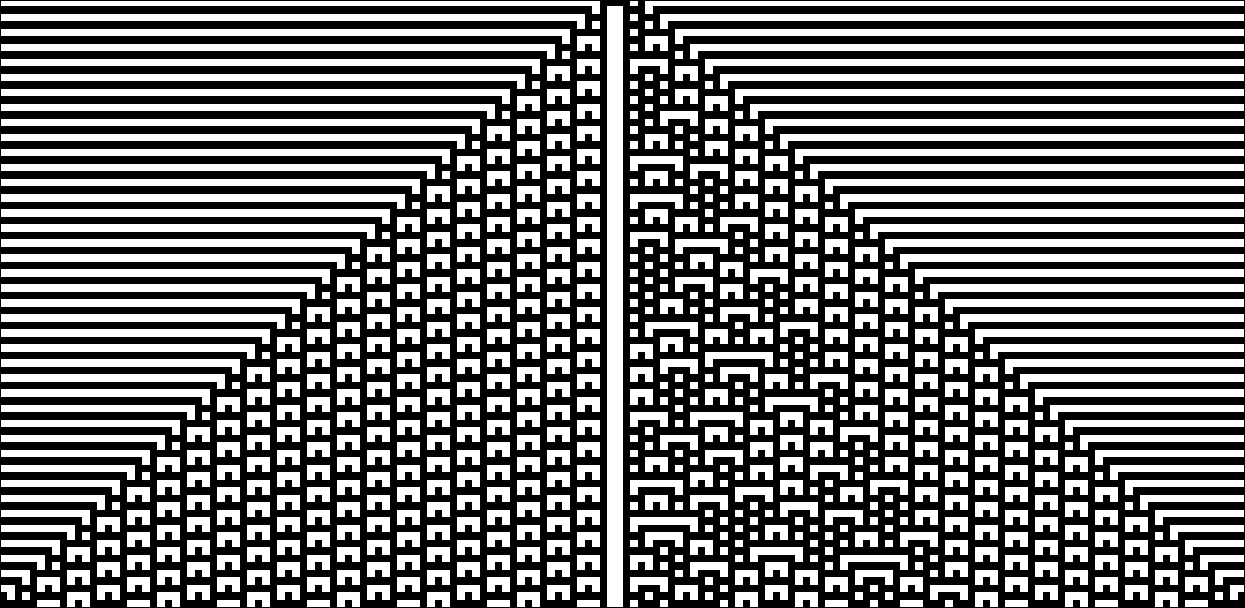}}}\end{center}
\caption{To which of Wolfram's Classes do these two ECAs (Rule 22 and Rule 109) belong? (a) Wolfram's ECA Rule 22 starting from a single black cell, (b) Rule 22 starting from another initial configuration (11001), (c) Wolfram's ECA Rule 109 starting from a single black cell, (d) The same Rule 109 starting from another initial configuration (111101).}
\label{zenilfig1}
\end{figure}

The motivation in \cite{kn:Zen10} is to address one of the apparent problems of Wolfram's original classification, that of rules behaving in different ways starting from different initial configurations. In the experiments that led Wolfram to propose his classification he started the systems with a ``random" initial configuration as a way to \textit{sample} the behaviour of a system and circumvent the problem of having to choose a particular initial configuration to map a system to its possible class of behaviour. The problem resides in the fact that a CA, like any other dynamical system, may have phase transitions, behaving very differently for different initial configurations (the question is ultimately undecidable as pointed out in \cite{kn:CY88}) but this is also a practical issue for an heuristic classification, given that systems may seem to jump from one class to another in such phase transitions. The chances of having a CA display an \textit{average} behaviour (that is, its behaviour for \textit{most} initial configurations) are greater when taking a ``random" initial configuration, only if one assumes that there is no bias towards any particular region of the possible enumerations of initial configurations (consider the behaviour of a CA starting from one initial configuration versus another (see Figures~\ref{zenilfig1}).\index{initial configurations}
\index{asymptotic behaviour}
In \cite{kn:Zen10} this issue is addressed with the definition of a compression-based phase transition coefficient capturing the \textit{asymptotic behaviour} of a system, which in turn allows to separate the collapsed classes and even advance a different and alternative classification, based on the sensitivity of a CA to its initial conditions, which has also been conjectured to be related to the system's ability to transfer information, and ultimately to its computing abilities, particularly as these relate to Turing universal computation (see \cite{kn:Zen12}). This approach does not solve the problem of a system that behaves in a  qualitatively different manner after a certain number of initial input configurations or after a certain period of time (the same problem encountered when devising the original classification), which is not a problem of method, but is instead related to the general problem of induction and of reachability (hence to undecidability in general). Nonetheless it does address the problem of a reasonable definition of the ``average behaviour" of a system (in this case a CA) under the same assumptions made for other enumerations (viz. that enumerations, especially natural ones, have no distinct regions where a system starts behaving in a completely different fashion, making it impossible to talk about the convergence in behaviour of a system). Wolfram's classes can once again be separated using the compression-based approach in combination with the following classification \cite{kn:ZenAISB}, derived from a phase transition coefficient presented in \cite{kn:Zen10}:

\begin{myitemize}
\item Class I: insensitivity to initial configurations, inability to transfer information other than isolated bits;
\item Class II: sensitivity to initial conditions, ability to transfer some information;
\item Class III: insensitivity to initial configurations, inability to transfer information, perhaps due to lack of (evident means of) control;
\item Class IV: sensitivity to initial conditions, ability to transfer some information.
\end{myitemize}

One can only understand how Classes I and III can now be together in this classification on the basis of the qualitative treatment explained above. In other words, when one changes the initial configuration of a system in either of these two classes (I and III) the system's behaviour remains the same (each evolution is equally compressible), and it is therefore considered unable to or inefficient at transferring information or programming a CA to perform (universal) computation. On the other hand, this suggests that classes II and IV may be better at transferring information, even if they may do so in different ways. This classification may tell us that some classes  are more sensitive to initial configurations.

Together, the compression-based classifications capturing different behaviours of the systems capture other intuitive notions that one would expect from Wolfram's original classification. The values for ECA calculated in \cite{kn:Zen10}  yielded results that also suggest that one may be able to relate these measures to universality through the definition of Class IV, as given above (see \cite{kn:Zen12}).
\index{phase transitions}\index{information transfer}

%%%%%%%%%%%%%%%%%%%%%
\section{Universal CA Class IV versus Class III}

\index{Culik II, K} Karel Culik II and Sheng Yu\index{Yu, S.} have demonstrated \cite{kn:CY88} that whether a CA belongs to Class IV is undecidable.  Nevertheless, some approximations have been developed, with interesting results. The use of genetic programming by Melanie Mitchell\index{Mitchell, M.}, Rajarshi Das, Peter Hraber, and James Crutchfield \cite{kn:MHC93, kn:DMC94} to obtain sets of rules with particles and computations is a case in point. As indeed is Emmanuel Sapin's\index{Sapin, E.} calculation of a non-additive binary universal 2D CA with a genetic algorithm, the {\it R rule} \cite{kn:SBC04, kn:SBC07}. However, the use of evolutionary techniques has been limited to a small portion of complex CA with few states and small configurations. Up to now, brute force programming has been necessary to obtain monsters of complex patterns in huge spaces, as Eppstein shows in \cite{kn:Epp02}.

\begin{figure}%[th]
\begin{center}
\subfigure[]{\scalebox{0.13}{\includegraphics{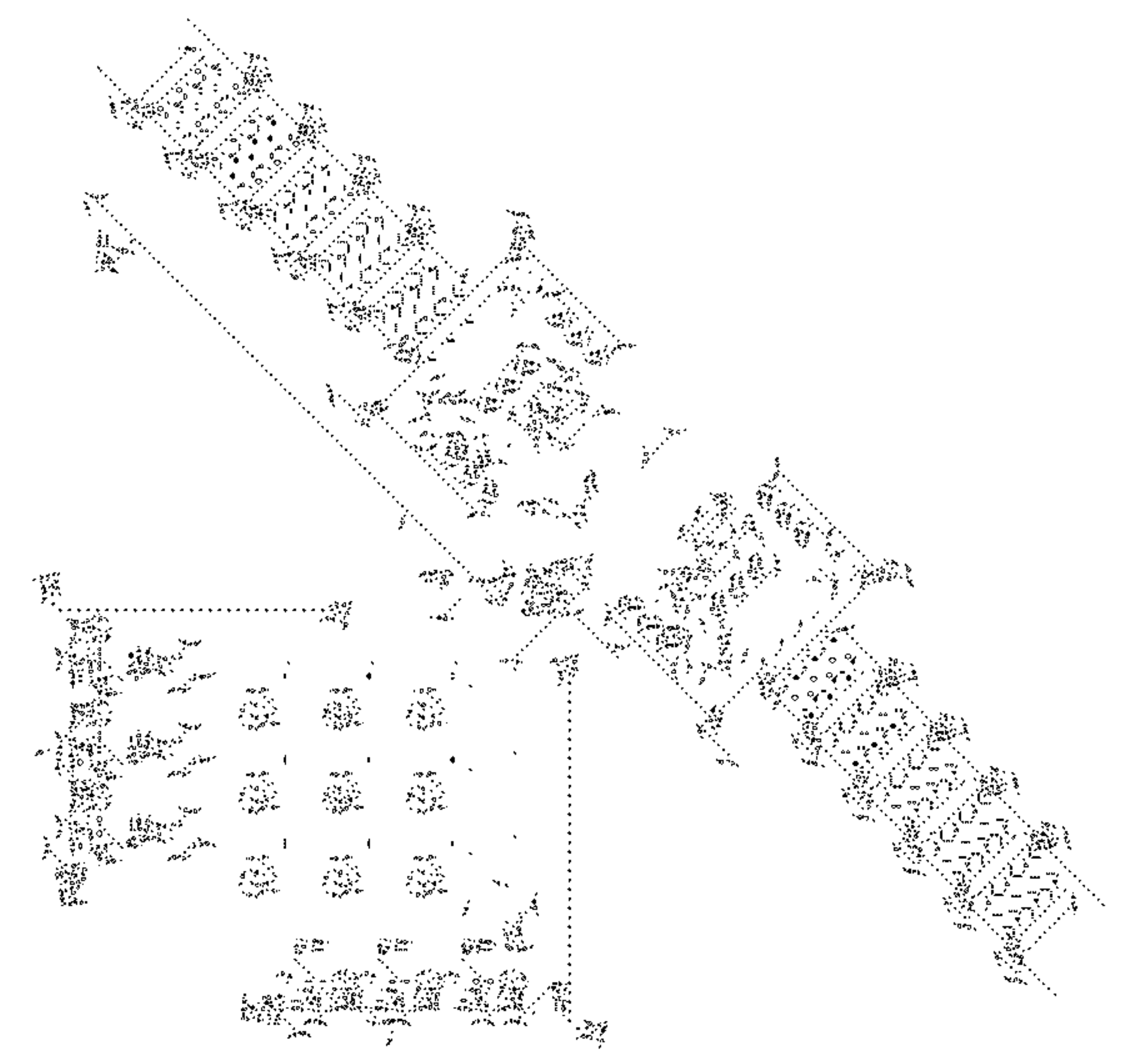}}} \hspace{0.8cm}
\subfigure[]{\scalebox{0.35}{\includegraphics{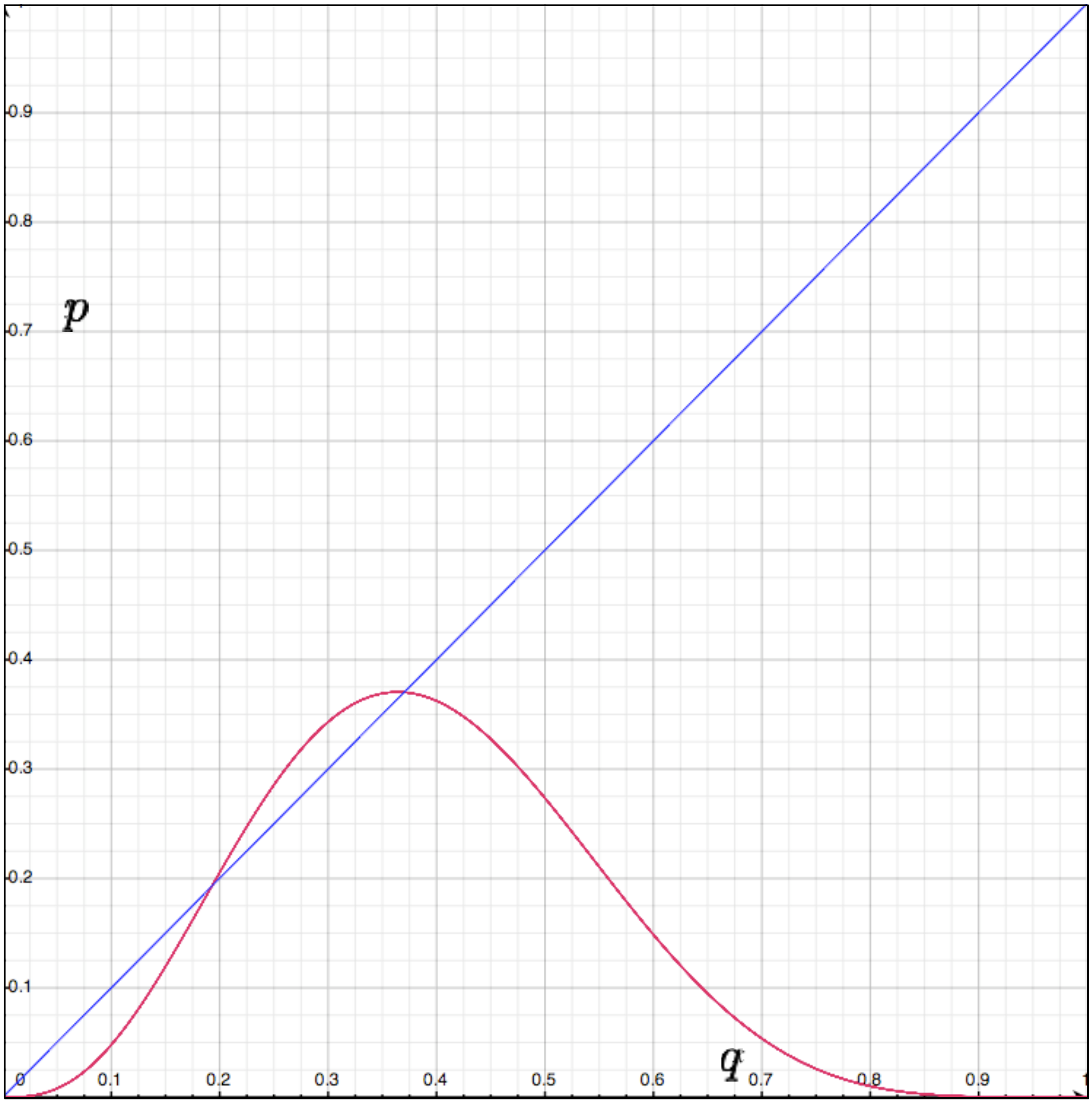}}}
\end{center}
\caption{(a) A 3-state, 3-symbol Turing machine in GoL by Rendell \cite{kn:Ren02, kn:Ren11}, (b) its mean field curve.}
\label{GoL_TM_MF}
\end{figure}

%%%%%%%%%%%%%%%%%%%%%
\subsection{The Game of Life: Class IV}
\index{Conway's Game of Life}
The most popular 2D CA is certainly Conway's Game of Life (GoL), a binary 2D additive CA, first published in Martin Garden's column in {\it Scientific American} \cite{kn:Gard70}. GoL can be represented as $R(2,3,3,3)$, or typically, as the  $B3/S23$ rule.\footnote{An excellent forum on GoL is ``LifeWiki'' \url{http://conwaylife.com/wiki/index.php?title=Main_Page}. To complement this, you may consult  ``The Game of Life Sites'' \url{http://uncomp.uwe.ac.uk/genaro/Cellular_Automata_Repository/Life.html}.} In 1982, Conway proved that GoL was universal by developing a register machine working with gliders, glider guns, still life and oscillator collisions \cite{kn:BCG82}. However, such universality was completed by Paul Rendell's demonstration in 2000 that involved implementing a 3-state, 3-symbol Turing machine in GoL \cite{kn:Ren02, kn:Ren11}. The machine duplicates a pattern of 1's within two 1's on the tape to the right of the reading position, running 16 cycles to stop with four 1's on the tape. A snapshot of this implementation is provided in Fig.~\ref{GoL_TM_MF}a. For details about each part and about the functionality of this machine please visit ``Details of a Turing Machine in Conway's Game of Life'' \url{http://rendell-attic.org/gol/tmdetails.htm}. 

GoL is a typical Class IV CA evolving with complex global and local behaviour. In its evolution space we can see a number of complex patterns which emerge from different configurations. GoL has been studied since 1969 by Conway\index{Conway, J.}, and William Gosper\index{Gosper, W.} of MIT's Artificial Life research group has taken a strong interest in it. The tradition of GoL research is very much alive, with today's GoL researchers discovering new and very complex constructions by running complicated algorithms. Just last year, GoL celebrated its 40th anniversary. The occasion was marked by the publication of the volume ``Game of Life Cellular Automata'' \cite{kn:Ada10}, summarising a number of contemporary and historical results in GoL research as well as work on other interesting Life-like rules. 

According to Mean field theory, $p$ is the probability of a cell's being in state `1' while $q$ is its probability of its being in state `0' i.e., $q=1-p$, and the {\it mean field equation} represents the neighbourhood that meets the requirement for a live cell in the next generation \cite{kn:Mc90}. As we have already seen, horizontal plus diagonal tangency, not crossing the identity axis (diagonal), and the marginal stability of the fixed point(s) due to their multiplicity indicates Wolfram's Class IV \cite{kn:Guto89}, or complex behaviour. Hence, we will review the global behaviour of GoL using Mean field theory. Figure~\ref{GoL_TM_MF}b shows the mean field curve for GoL, with polynomial:

$$p_{t+1}=28p^3_tq^5_t(2p_t+3q_t).$$

The origin is a stable fixed point, while the unstable fixed point $p=0.2$ represents the fact that densities around 20\% induce complex behaviour for configurations in such a distribution. $p=0.37$ is the maximum stable fixed point where GoL commonly reaches global stability inside the evolution space.

In \cite{kn:ZenJETAI} a compression-based phase transition coefficient was calculated, showing that, as expected, GoL exhibits a high degree of variability and potential (efficient) programmability. This is in agreement with the known fact that GoL is capable of universal computation, and hence supports the idea that sensitivity to initial configurations is deeply connected to both programmability and (Turing) universality.

\begin{figure}[th]
\centering
\includegraphics[width=0.6\textwidth]{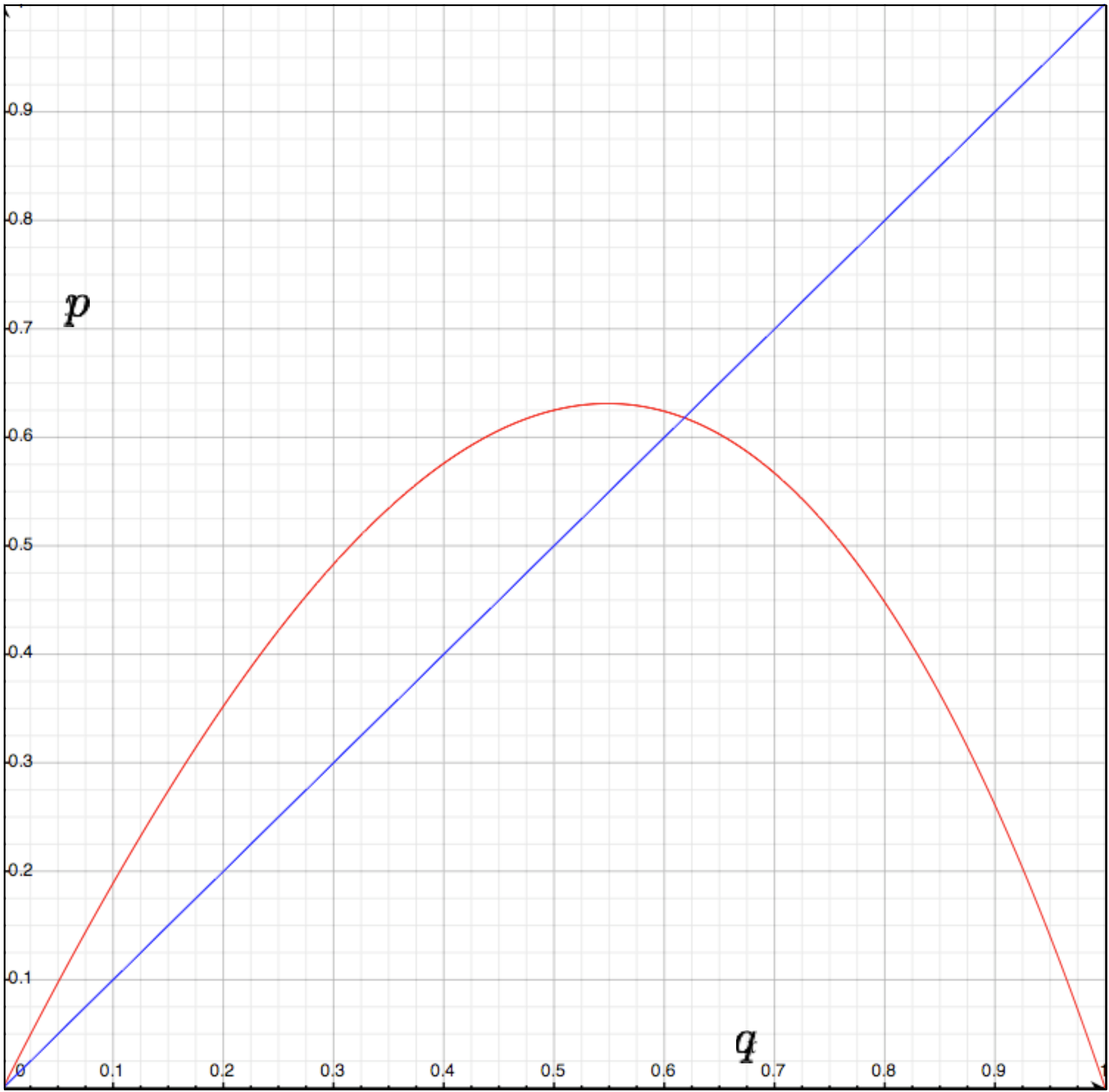}
\caption{Mean field curve for ECA Rule 110.}
\label{Rule110_MF}
\end{figure}

 \begin{figure}%[th]
\begin{center}
\subfigure[]{\scalebox{0.22}{\includegraphics{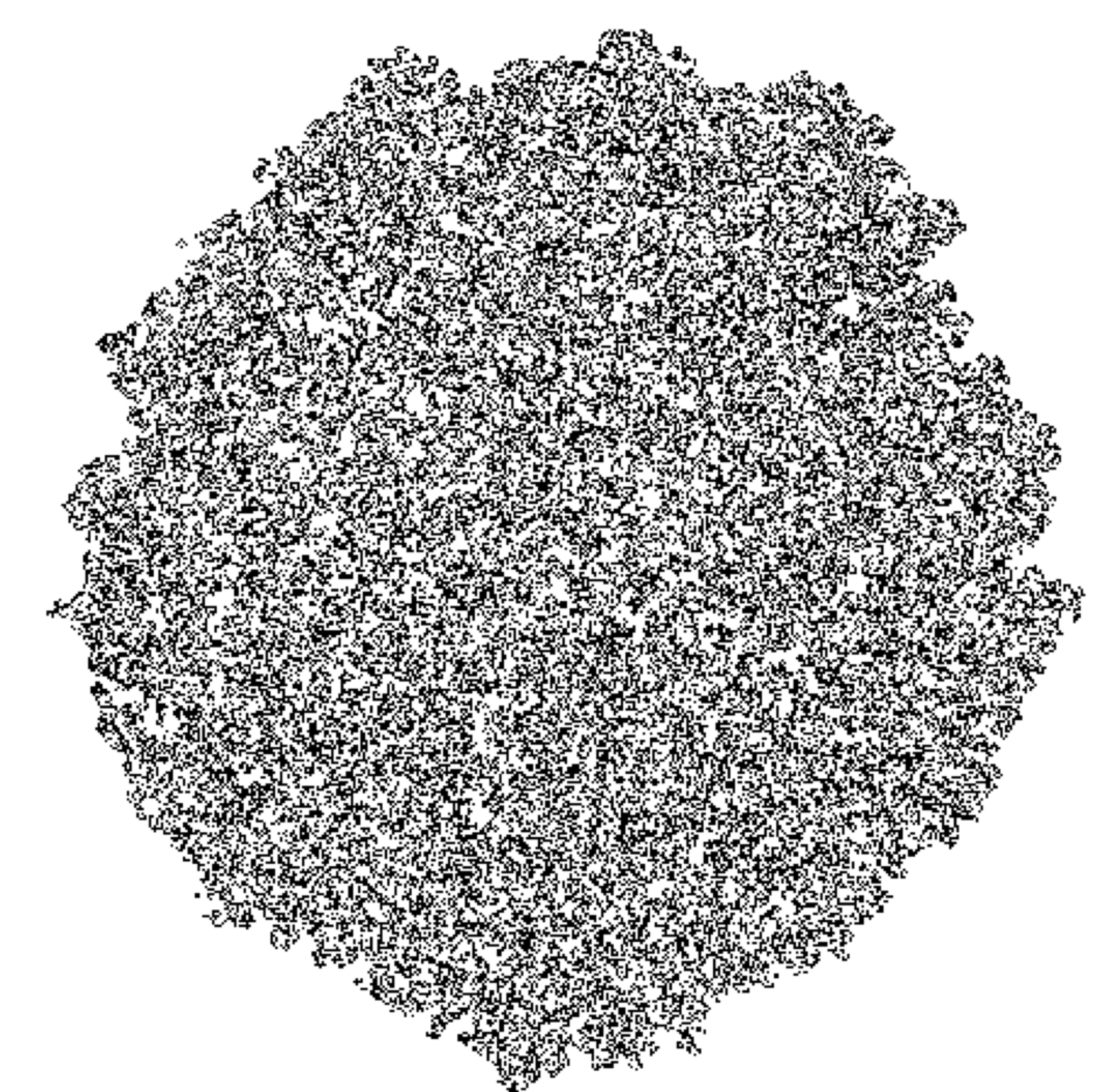}}} \hspace{0.8cm}
\subfigure[]{\scalebox{0.34}{\includegraphics{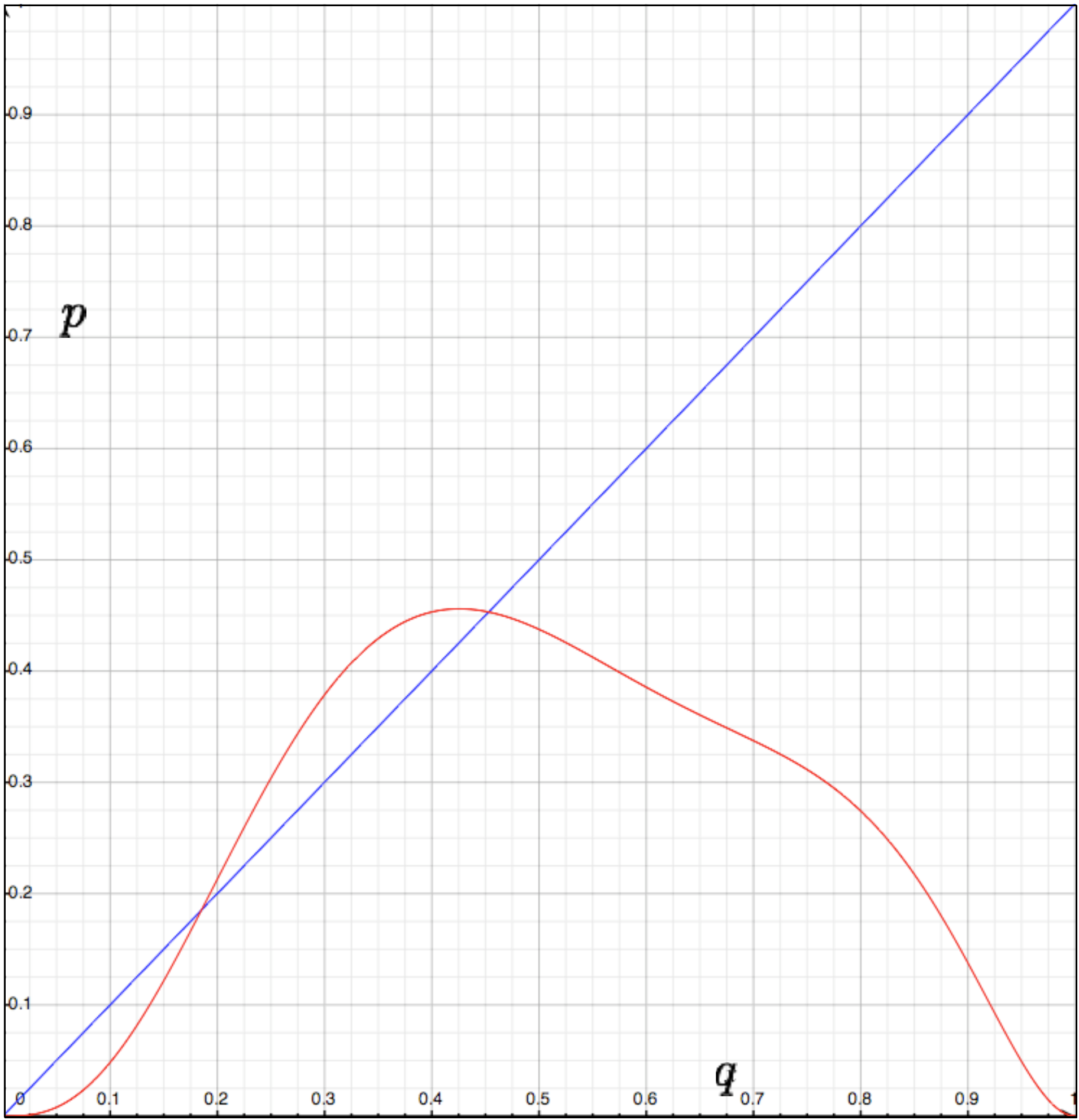}}}
\end{center}
\caption{(a) Evolution starting from an L-pentomino in Life-like CA $B35/S236$, (b) its mean field curve.}
\label{B35_S236}
\end{figure}

%%%%%%%%%%%%%%%%%%%%%
\subsection{Life-like rule $B35/S236$: Class III}

The Life-like CA evolution rule $B35/S236$ was proposed by Eppstein and Dean Hickerson as a chaotic CA with sufficient elements for developing universality. Details about these computable elements are available at \url{http://www.ics.uci.edu/~eppstein/ca/b35s236/construct.html}. The family of gliders and other complex constructions in this rule can be found at \url{http://www.ics.uci.edu/~eppstein/ca/b35s236/}.

 The $B35/S236$ automaton commonly evolves chaotically. Figure~\ref{B35_S236}a displays a typical chaotic evolution starting from an L-pentomino configuration; after 1,497 generations  there is a population of 52,619 live cells. Here we see how a few gliders emerge from chaos and then quickly escape, although the predominant evolution over a long period is chaotic.

 Figure~\ref{B35_S236}b shows the mean field curve for CA $B35/S236$, with polynomial:

$$
p_{t+1}=28p^3_tp^2_t(p^4_t+2p_tq^3_t+2p^2_tq^2_t+3q^4_t).
$$

The origin is a stable fixed point (as in GoL) which guarantees the stable configuration in zero, while the unstable fixed point $p=0.1943$  (again very similar to GoL) represents densities where we could find complex patterns emerging in $B35/S236$. $p=0.4537$ is the maximum stable fixed point at which $B35/S236$ commonly reaches global stability.

 This way, $B35/S236$ preserves the diagonal tangency between a stable and an unstable fixed point on its mean field curve. But although its values are close to those of GoL, CA $B35/S236$ has a bigger population of live cells, which is not a sufficient condition for constructing reliable organisms from unreliable components.

%%%%%%%%%%%%%%%%%%%%%
\subsection{ECA Rule 110: Class IV}
\label{rule110}

The 1D binary CA rule numbered 110 in Wolfram's system of classification \cite{kn:Wolf83} has been the object of special attention due to the structures or gliders which have been observed in instances of its evolution from random initial conditions. The rule is assigned number 110 in Wolfram's enumeration because it represents the decimal base of the transition rule expanded in binary: 01110110. The transition function evaluates the neighbourhoods synchronously in order to calculate the new configuration transforming the neighbourhoods 001, 010, 011, 101 and 011 into state 1 and the neighbourhoods 000, 100 and 111 into state 0. It has been suggested that Rule 110 belongs to the exceptional Class IV of automata whose chaotic aspects are mixed with regular patterns. But in this case the background where the chaotic behaviour occurs is textured rather than quiescent, a tacit assumption in the original classification.\footnote{A repository of materials on ECA Rule 110 can be found at: \url{http://uncomp.uwe.ac.uk/genaro/Rule110.html}.} Rule 110 was granted its own appendix (Table 15) in \cite{kn:Wolf86}. It contains specimens of evolution including a list of thirteen gliders compiled by Lind and also presents the conjecture that the rule could be universal. 

The literature on the origins of Rule 110 includes a statistical study done by Wentian Li and Mats Nordahl in 1992 \cite{kn:LN92}. This paper studies the transitional role of Rule 110 and its relation to Class IV rules figuring between Wolfram's classes II and III. The study would seem to reflect an approach to equilibrium statistics via a power law rather than exponentially.
\index{Cook, M.}
Matthew Cook wrote an eight page introduction \cite{kn:Cook99} listing gliders from $A$ through $H$ and a glider gun.\footnote{An extended list of gliders in Rule 110 is provided in \url{http://uncomp.uwe.ac.uk/genaro/rule110/glidersRule110.html}.}. This list shows new gliders which do not appear on Lind's list, gliders with rare extensions, and a pair of gliders of complicated construction, including an amazing glider gun. Cook makes a comparison between Rule 110 and Life, finding some similarities in the behaviour of the two evolution rules and suggesting that Rule 110 may be called ``LeftLife.''

Looking at the rule itself, one notices a ubiquitous background texture which Cook calls ``ether,'' although it is just one of many regular stable lattices capable of being formed by the evolution rule, and can be obtained quickly using the de Bruijn diagrams \cite{kn:Mc99, kn:MMS06}. \index{de Bruijn diagram}
\index{McIntosh, H.V.}
McIntosh raises the issue of the triangles of different sizes that cover the evolution space of Rule 110 \cite{kn:Mc00}. The appearance of these triangles suggests the analysis of the plane generated by the evolution of Rule 110 as a two dimensional shift of finite type. This suggestion is arrived at by observing that the basic entities in the lattices, the unit cells, induce the formation of upside-down isosceles right triangles of varying sizes. The significance of Rule 110 could lie in the fact that it is assembled from recognisably distinct tiles, and hence its evolution can be studied as a tiling problem, in the sense of Hao Wang \cite{kn:GS82}. It may even be possible to see fitting elements of one lattice into another as an instance of Emil L. Post's correspondence principle \cite{kn:Dav94}, which would establish the computational complexity of the evolution rule \cite{kn:Mc99}.

The most important result both in the study of Rule 110 and in CA theory over the last twenty years, is the demonstration that Rule 110 is capable of universal computation \cite{kn:Cook04, kn:Wolf02, kn:Mc02, kn:Cook11, kn:MMS11}. For so a type of system called a cyclic tag system (CTS) as a variation of a well-known model of computation (Post's tag systems) was designed to be of use for the proof and its characteristic restrictions: 1D, boundary conditions, package of gliders, and multiple collisions. CTS are a new kind of computing formalism \cite{kn:Cook04, kn:Wolf02} used as tools for implementing computations in Rule 110.

Fig.~\ref{Rule110_MF}b shows the mean field curve for Rule 110 with polynomial:

$$
p_{t+1} = 2p_tq^{2}_t+3p^{2}_tq_t.
$$

The origin of Fig.~\ref{Rule110_MF} displays a stable fixed point (as in GoL) which guarantees the stable configuration in zero. The maximum point ($p=0.6311$) is close to the fixed stable point in $p=0.62$. In Rule 110 we cannot find unstable fixed points, and in any case the emergence of complex structures is ample and diverse.

A basin (of attraction) field of a finite CA is the set of basins of attraction into which all possible states and trajectories will be organised by the local function $\varphi$. The topology of a single basin of attraction may be represented by a diagram, the {\it state transition graph}. Thus the set of graphs composing the field specifies the global behaviour of the system \cite{kn:WL92}.

As calculated in \cite{kn:Zen10}, rules such as Rule 110 and Rule 54 (also believed to be capable of universal computation) had a large compression-based phase transition coefficient, as discussed in Section \ref{compressibility}, meaning that their ability to transfer information was well captured by the measure defined in \cite{kn:Zen10} (and, interestingly, perhaps strengthens the belief that Rule 54 is capable of Turing universality).

\section{\textit{Heat} and \textit{programmability} in Class III}
\label{heat}

Class III CAs may turn out to be too sensitive, so the question may be whether even if they are that sensitive they can carry information from one side to another. Universality results in simple programs capable of complicated behaviour have traditionally relied on localised structures (``particles") well separated by relatively uniform regions. It could also be the case that proofs of universality of seemingly Class IV systems are easier to construct because of its ``particle-like" behaviour, unlike systems seemingly in Class III.

The open problem is thus to prove computational universality in a simple program system for which an entropy measure on each time step remains near its maximum -- e.g. 80\% of its maximum theoretical value on at least 80\% of its time steps.  Can a ``hot system" of this sort perform meaningful computation?

In the Game of Life, for example, there is a common intuitive notion of \textit{heat}\footnote{See \url{http://www.argentum.freeserve.co.uk/life.htm} accessed in July 2012.}, defined as the average number of cells which change state in each generation (note the strong connection of Shannon's Entropy and the Mean Field Theory). For example, \textit{the heat} of a glider in GoL is known to be four, because two cells are born and two die in every generation, and that for a blinker is 4, because 2 cells are born and 2 die in every generation. In general, for a period $n$ oscillator with an $r$-cell ``rotor", the heat is at least $2r/n$, and no more than $r(1-(n \mod 2)/n)$. 

The concept of \textit{heat} can clearly be associated with Wolfram's chaotic Class III, where CAs, e.g., rule 30, change state at a very high rate, (see Figures (c)~\ref{WolframClasses}), which is what keeps them from developing persistent structures such as are seen in Rule 110 (see Figure (d)~\ref{WolframClasses}). The presence of persistent structures in Wolfram's Rule 110 and Conway's Game of Life is what traditionally has been used to perform computation--implementing logic gates or transferring information over time by putting particles in the way of interacting with each other. So the question is whether CAs such as the ones belonging to Wolfram's Class III are ``too hot" to transfer information and are therefore, paradoxically in this particular way, just like Class I systems which are unable to perform computation. Alternatively, Class III may be able to perform computation, as has been suggested, but it may turn out to be difficult to program such systems (if not designed to ``look" like a Class III system by using first a system from a Class IV, somehow \textit{hiding} the computing capabilities of the Class III system), and this potential similarity between the insensitivity to initial conditions of Class I and Class III systems is what the compressibility approach discussed in Section \ref{compressibility} is measuring and which has been advanced in \cite{kn:ZenAISB} as a measure of \textit{programmability}.
\index{entropy}\index{heat}\index{dynamic systems}\index{programmability}\index{Kolmogorov complexity}\index{enumeration}\index{cellular automata}
Wolfram identified some of these issues in his enumeration of open problems in the research on CA \cite{wolframopen} (problems 1, 2 and 14), concerning the connections between the computational and statistical characteristics of cellular automata, measures of entropy and complexity and how to improve his classification using dynamic systems (which was one of the motivations of \cite{kn:Zen10}). Wolfram asks, for example, about the rate of information transmission of a CA in relation to its \index{Lyapunov exponents}Lyapunov exponent (positive for Classes III and IV) and the computational power of these systems according to their classes. Another interesting question concerns the connection to Langton's $\lambda$ parameter \cite{kn:Lang84} and the ongoing investigation of its connections to some of the approaches described in this paper. In \cite{baetens} a similar approach is taken using Lyapunov exponents and Jacobians-- anticipated by Wolfram in \cite{kn:Wolf86}--where the calculation of the number of cells that differ provide a metric of the average rate of transmission of information (one that is related to the more informal term \textit{heat} in GoL).
\index{Langton, C.}\index{information transfer}

\section{Final remarks}
\label{finalremarks}

Usually, Class III rules are not considered candidates for computational universality. However, in some cases  such rules can support complex patterns, including performing complex computations. Exploring many CA rules, including the exceptionally chaotic Life-like rule {\it Dead without Life} \cite{kn:GM96}, one finds that there are several rules between chaos and complexity which are not included within the domain of complex behaviour. However, they present many elements equally likely to reach Turing computational universality. An important point made in this survey and review is that it seems clearly to be the case that it is not only {\it complex CA}\footnote{A Complex Cellular Automata Repository with several interesting rules is available at \url{http://uncomp.uwe.ac.uk/genaro/otherRules.html}. We particularly recommend Tim Hutton's {\it Rule Table Repository} \url{http://code.google.com/p/ruletablerepository/}.} rules that are capable of computation, and that CA, even if simple or random-looking, may support Turing universality. Whether the encoding to make them actually compute turns out to be more difficult than taking advantage of the common interacting persistent structures in rules usually believed to belong to Wolfram's class IV is an open question. 

Previous results on universal CAs (developing signals, self-reproductions, gliders, collisions, tiles, leaders, etc.) prove that unconventional computing can be obtained depending on the nature of each complex system. For example, to prove universality in Rule 110 it was necessary to develop a new equivalent Turing machine to take advantage of limitations in 1D and the same dynamics in its evolution space, e.g., mobility of gliders and boundary properties. Hence, a CTS was devised, before this system was known as a circular machine \cite{kn:Arb69, kn:KR01, kn:Mor07, kn:MAS11}. This way, the nature of each system would determine the best environment in which to design a corresponding computer. This could be the basis of Wolfram's {\it Principle of Computational Equivalence} and it is also the inspiration behind the definition of \textit{programmability} measures for natural computation in \cite{kn:ZenAISB}. Wolfram's {\it Principle of Computational Equivalence} ultimately only distinguishes between two kinds of behaviours (despite Wolfram's own heuristic classification), namely those that are ``sophisticated" enough and reach  \textit{Wolfram's threshold}, constituting a class of systems capable of computational universality, and those that fall below this threshold and are incapable of universal computation. And indeed, the compression-based classification in \cite{kn:Zen10} at first distinguishes only two classes. 

A number of approximations were developed or adapted to find complex CA. Perhaps the most successful technique was the one developed by Wuensche, with its $Z$ parameter \cite{kn:Wue99}. Some attempts were made by Mitchell {\it et. al} using genetic algorithms, although they had a particular interest in finding rules able to support complex patterns (gliders) with computational uses \cite{kn:DMC94, kn:WO08}. Unfortunately, these algorithms have strong limitations when it comes to searching in large rule spaces and very complex structures. And though the technique in \cite{kn:Zen10} has proven capable of identifying complex systems with great accuracy, it requires very large computational resources to extend the method to larger rule spaces if a thorough investigation is desired (though in conjunction with other techniques it may turn out to be feasible).

 As it has proven to be a very rich space, new kinds of CAs are proposed all the time. e.g., reversible CA \cite{kn:Kari96, kn:SHM08, kn:Mc91a}, partitioned CA \cite{kn:Wolf02}, hyperbolic CA \cite{kn:Mar07}, CA with non-trivial collective behaviour (self-organization) \cite{kn:CM92, kn:CGG08}, asynchronous CA \cite{kn:FM05}, biodiversity in CA \cite{kn:MAM}, CA with memory \cite{kn:Alo09, kn:Alo11}, morphological diversity \cite{kn:AM10}, identification of CA \cite{kn:Ada94}, communication complexity \cite{kn:DRT04, kn:GMR11}, pattern recognition from CA \cite{kn:AN86}, to mention a few. 

Some other studies dedicated to designing or identifying universal CAs are \cite{kn:Hey98, kn:Ada01, kn:Ada02, kn:GM03, kn:MAS10}. Obtaining CA of Class IV from other rules has been studied via lattice analysis \cite{kn:Gunji10}, with memory \cite{kn:MAA10, kn:MAA12, kn:MAS10, kn:Alo03, kn:AM03, kn:Alo06}, asynchronous \cite{kn:Suzu94, kn:TV02, kn:CBM02, kn:FM05}, differential equations \cite{kn:Chua06}, partitioned \cite{kn:MH89, kn:Mor90, kn:IM00, kn:Mor07, kn:Mor08, kn:MMI99, kn:Marg98, kn:Marg03}, parity-filter CA \cite{kn:PST86, kn:Siw02, kn:JSS01}, number-conserving \cite{kn:MTI02} changing different neighbourhoods in CA \cite{kn:WN09}.

 CA as {\it super computer models} are developed extensively in \cite{kn:von66, kn:Codd68, kn:Ban71, kn:Marg84, kn:MTV86, kn:TM87, kn:Wolf88, kn:Sipp97, kn:Hey98, kn:Toff98, kn:FT01, kn:Ada01, kn:Ada02, kn:Ada02a, kn:ACA05, kn:Wors09, kn:Hutt10, kn:MAS11}.

%%%%%%%
\section*{Acknowledgements}
G. J. Mart{\'i}nez wants to thank support given by EPSRC grant EP/F054343/1, J. C. Seck-Tuoh-Mora wants to thank support provided by CONACYT project CB-2007-83554 and H. Zenil wants to thank support by the FQXi under grant number FQXi-MGA-1212.

%%%%%%%%%%%%%%%%%%%%%


\begin{thebibliography}{999}

\bibitem{kn:ACA05} Adamatzky, A., Costello, B. L., \& Asai, T. (2005) \textit{Reaction-Diffusion Computers}, Elsevier.

\bibitem{kn:Ada94} Adamatzky, A. (1994) \textit{Identification of Cellular Automata}, Taylor \& Francis.

\bibitem{kn:Ada01} Adamatzky, A. (2001) \textit{Computing in Nonlinear Media and Automata Collectives}, Institute of Physics Publishing, Bristol and Philadelphia.

\bibitem{kn:Ada02} Adamatzky, A. (Ed.) (2002) \textit{Collision-Based Computing}, Springer.

\bibitem{kn:Ada02a} Adamatzky, A. (2002) New media for collision-based computing, In: \textit{Collision-Based Computing}, A. Adamatzky (Ed.), Springer, 411--442.

\bibitem{kn:Ada10} Adamatzky, A. (Ed.) (2010) \textit{Game of Life Cellular Automata}, Springer.

\bibitem{kn:Alo03} Alonso-Sanz, R. (2003) Reversible Cellular Automata with Memory, \textit{Physica D} {\bf 175} 1--30.

\bibitem{kn:Alo06} Alonso-Sanz, R. (2006) Elementary rules with elementary memory rules: the case of linear rules, \textit{Journal of Cellular Automata} {\bf 1} 71--87.

\bibitem{kn:Alo09} Alonso-Sanz, R. (2009) \textit{Cellular Automata with Memory}, Old City Publishing.

\bibitem{kn:Alo11} Alonso-Sanz, R. (2011) \textit{Discrete Systems with Memory}, World Scientific Series on Nonlinear Science, Series A.

\bibitem{kn:AM03} Alonso-Sanz, R. \& Martin, M. (2003) Elementary CA with memory, \textit{Complex Systems} {\bf 14(2)} 99--126.

\bibitem{kn:AM10} Adamatzky, A. \& Mart{\'i}nez, G. J. (2010) On generative morphological diversity of elementary cellular automata,'' \textit{Kybernetes} {\bf 39(1)} 72--82.

\bibitem{kn:AN86} Aizawa, Y. \& Nishikawa, I. (1986) Toward the classification of the patterns generated by one-dimensional cellular automata, In: \textit{Dynamical Systems and Nonlinear Oscillators}, Giko Ikegami (Ed.), World Scientific Press, 210-222.

\bibitem{kn:Arb69} Arbib, M. A. (1969) Monogenic normal systems are universal, \textit{Journal of the Australian Mathematical} {\bf 3(3)} 301--306.

\bibitem{baetens} J.M. Baetens \& B. De Baets (2010). Phenomenological study of irregular cellular automata based on Lyapunov exponents and Jacobians, \textit{Chaos}, 20.

\bibitem{kn:Ban71} Banks, E. R. (1971) Information and transmission in cellular automata, \textit{PhD Dissertion}, Cambridge, MA, MIT.

\bibitem{kn:BCG82} Berlekamp,~E. R., Conway,~J. H., \& Guy,~R. K. (1982) \textit{Winning Ways for your Mathematical Plays}, Academic Press, (vol. 2, chapter 25).

%\bibitem{kn:BNR91} Boccara, N., Nasser, J., \& Roger, M. (1991) Particle like structures and their interactions in spatio-temporal patterns generated by one-dimensional deterministic cellular automaton rules, \textit{Physical Review A} {\bf 44(2)} 866--875.

\bibitem{kn:CBM02} Clapham, N. T., Barlow, M., McKay, R. I. (2002) Toward classifying randomly asynchronous cellular automata, In: \textit{Proceedings of the 6th International Conference on Complex Systems (CS02)}, 63--71

%\bibitem{kn:Chap02} Chapman, P. (2002) Life Universal Computer, \url{http://www.igblan.free-online.co.uk/igblan/ca/}.

\bibitem{kn:Chua06} Chua, L. (2006--12) \textit{A Nonlinear Dynamics Perspective of Wolfram's New Kind of Science, Vol. 1, 2, 3, 4, 5}, World Scientific Publishing Company.

\bibitem{kn:Codd68} Codd, E. F. (1968) \textit{Cellular Automata}, Academic Press, Inc. New York and London.

\bibitem{kn:Cook99} Cook, M. (1998) Introduction to the activity of Rule 110 (copyright 1994-1998 Matthew Cook), \url{http://w3.datanet.hu/~cook/Workshop/CellAut/Elementary/Rule110/}.

\bibitem{kn:Cook04} Cook, M. (2004) Universality in Elementary Cellular Automata, \textit{Complex Systems} {\bf 15(1)} 1--40.

\bibitem{kn:Cook11} Cook, M. (2011) A Concrete View of Rule 110 Computation, In: \textit{The Complexity of Simple Programs} T. Neary, D. Woods, A. K. Seda, \& N. Murphy (Eds.), 31--55, Elsevier.

\bibitem{kn:CM92} Chat\'e, H. \& Manneville, P. (1992) Collective behaviours in Spatially Extended Systems with Local Interactions and Synchronous Updating, \textit{Progress in Theoretical Physics} {\bf 87} 1--60.

\bibitem{kn:CGG08} Chat\'e, H., Ginelli, F., Gr\'egoire, G., Peruani, F. \& Raynaud, F. (2008) Modeling collective motion: variations on the Vicsek model, \textit{The European Physical Journal B} {\bf 64(3-4)} 451--456.

\bibitem{kn:CY88} Culik II, K. \& Yu, S. (1988) Undecidability of CA Classification Schemes, \textit{Complex Systems} {\bf 2(2)} 177--190.

\bibitem{kn:Dav94} Davis, M. (Ed.) (1994) \textit{Solvability, Provability, Definability: The Collected Works of Emil L. Post}, Birkh\"auser Boston.

\bibitem{kn:DMC94} Das, R., Mitchell, M., \& Crutchfield, J. P. (1994) A genetic algorithm discovers particle-based computation in cellular automata, In: \textit{Parallel Problem Solving from Nature---PPSN III}, Y. Davidor, H.-P. Schwefel, \& R. Manner (Eds.) 344--353, Springer-Verlag London.

\bibitem{kn:DRT04} Durr, C., Rapaport, I., \& Theyssier, G. (2004) Cellular automata and communication complexity, \textit{Theoretical Computer Science} {\bf 322} 355--368.

\bibitem{kn:Epp99} Eppstein, D. (1999) Wolfram's Classification of Cellular Automata, \url{http://www.ics.uci.edu/~eppstein/ca/wolfram.html}.

\bibitem{kn:Epp02} Eppstein, D. (2002) Searching for spaceships, \textit{MSRI Publications} {\bf 42} 433--452.

\bibitem{kn:FM05} Fat\`{e}s, N. \& Morvan, M. (2005) An Experimental Study of Robustness to Asynchronism for Elementary Cellular Automata, \textit{Complex Systems} {\bf 16} 1--27.

\bibitem{kn:FT01} Fredkin, E. \& Toffoli, T. (2001) Design Principles for Achieving High-Performance Submicron Digital Technologies, In: \textit{Game of Life Cellular Automata}, A. Adamatzky (Ed.), Springer, 27--46.

\bibitem{kn:Gard70} Gardner, M. (1970) Mathematical Games --- The fantastic combinations of John H. Conway's new solitaire game Life, \textit{Scientific American} {\bf 223} 120--123.

\bibitem{kn:GM96} Griffeath, D. \& Moore, C. (1996) Life Without Death is P-complete, \textit{Complex Systems} {\bf 10(6)} 437--447.

\bibitem{kn:GM03} Griffeath, D. \& Moore, C. (Eds.) (2003) \textit{New constructions in cellular automata}, Oxford University Press.

\bibitem{kn:GS82} Gr\"{u}nbaum, B. \& Shephard, G. C. (1987) \textit{Tilings and Patterns}, W. H. Freeman and Company, New York.

\bibitem{kn:GMR11} Goles, E., Moreira, A., \& Rapaport, I. (2011) Communication complexity in number-conserving and monotone cellular automata, \textit{Theoretical Computer Science} {\bf 412} 3616--3628.

%\bibitem{kn:Gou09} Goucher, A. P. (2009) Completed Universal Computer/Constructor, In: Pentadecathlon website \url{http://pentadecathlon.com/lifeNews/2009/08/post.html}.

%\bibitem{kn:Gou10} Goucher, A. P. (2010) Universal Computation in GoL Cellular Automata, In: \textit{Game of Life Cellular Automata}, A. Adamatzky (Ed.), Springer, 505--518.

\bibitem{kn:Gunji10} Gunji, Y-P (2010) Inducing Class 4 behaviour on the Basis of Lattice Analysis, \textit{Complex Systems} {\bf 19(5)} 177--194.

\bibitem{kn:GVK87} Gutowitz, H. A., Victor, J. D., \& Knight, B. W. (1987) Local structure theory for cellular automata, \textit{Physica D} {\bf 28} 18--48.

\bibitem{kn:Guto87a} Gutowitz, H. A. \& Victor, J. D. (1987) Local structure theory in more that one dimension, \textit{Complex Systems} {\bf 1(1)} 57--68.

\bibitem{kn:Guto89} Gutowitz H. A. (1989) Mean Field vs. Wolfram Classification of Cellular Automata. Historical link \url{http://www.santafe.edu/~hag/mfw/mfw.html}. Functional link \url{http://citeseerx.ist.psu.edu/viewdoc/summary?doi=10.1.1.29.4525}.

\bibitem{kn:Guto89a} Gutowitz, H. A. \& Victor, J. D. (1999) Local structure theory: calculation on hexagonal arrays, and iteraction of rule and interaction of rule and lattice, \textit{Journal of Statistical Physical} {\bf 54}  495--514.

\bibitem{kn:Hey98} Hey, A. J. G. (1998) \textit{Feynman and computation: exploring the limits of computers}, Perseus Books.

%\bibitem{kn:HC97} Hanson, J. E. \& Crutchfield, J. P. (1997) Computacional Mechanics of Cellular Automata: An Example, \textit{Physics D} {\bf 103} 169--189.

\bibitem{kn:Hutt10} Hutton, T. J. (2010) Codd's self-replicating computer, \textit{Artificial Life} {\bf 16(2)} 99--117.

\bibitem{kn:IM00} Imai, K. \& Morita, K. (2000) A computation-universal two-dimensional 8-state triangular reversible cellular automaton, \textit{Theoret. Comput. Sci.} {\bf 231} 181--191.

\bibitem{kn:JSS01} Jakubowski, M. H., Steiglitz, K., \& Squier, R. (2001) Computing with Solitons: A Review and Prospectus, \textit{Multiple-Valued Logic} {\bf 6(5-6)} (also republished in \cite{kn:Ada02}).

\bibitem{kn:Kari96} Kari, J. (1996) Representation of Reversible Cellular Automata with Block Permutations, \textit{Mathematical Systems Theory} {\bf 29(1)} 47-61.

\bibitem{kn:KR01} Kudlek, M. \& Rogozhin, Y. (2001) Small Universal Circular Post Machine, \textit{Computer Science Journal of Moldova} {\bf 9(25)} 34--52.

\bibitem{kn:Kau93} Kauffman, S. (1993) \textit{The Origins of Order: Self-Organization and Selection in Evolution}, Oxford University Press.

\bibitem{kn:Lang84} Langton, C. G. (1984) Self Reproduction in Cellular Automata, \textit{Physica D} {\bf 10} 135--144.

\bibitem{kn:Lang86} Langton, C. G. (1986) Studying Artificial Life with Cellular Automata, \textit{Physica D} {\bf 22(1-3)} 120--149.

%\bibitem{kn:Lang91} Langton, C. G. (1991) Life at the Edge of Chaos, \textit{Artificial Life II}, Addison-Wesley.

\bibitem{kn:LN92} Li, W \& Nordahl, M. G. (1992) Transient behaviour of cellular automaton Rule 110, \textit{Physics Letters A} {\bf 166} 335--339.

\bibitem{kn:LP90} Li, W. \& Packard, N. (1990) The Structure of the Elementary Cellular Automata Rule Space, \textit{Complex Systems} {\bf 4(3)} 281--297.

\bibitem{kn:MAM} Redeker, M., Adamatzky, A., \& Mart{\'i}nez, G. J., Expressiveness of Elementary Cellular Automata, {\it submitted}.

\bibitem{kn:MAA10} Mart{\'i}nez, G. J., Adamatzky, A., Alonso-Sanz, A., \& Seck-Tuoh-Mora, J. C. (2010) Complex dynamic emerging in Rule 30 with majority memory, \textit{Complex Systems} {\bf 18(3)} 345--365.

\bibitem{kn:MAA12} Mart{\'i}nez, G. J., Adamatzky, A., \& Alonso-Sanz, A., Complex dynamics of cellular automata emerging in chaotic rules, \textit{Int. J. Bifurcation and Chaos} {\bf 22(2)} 2012.

%\bibitem{kn:MAC08} Mart{\'i}nez, G. J., Adamatzky, A., \& Costello, B. C. L. (2008) On logical gates in precipitating medium: cellular automaton model, \textit{Physics Letters A} {\bf 1(48)} 1--5.

%\bibitem{kn:MAM06} Mart{\'i}nez, G. J., Adamatzky, A., \& McIntosh, H. V. (2006) Phenomenology of glider collisions in cellular automaton Rule 54 and associated logical gates, \textit{Chaos, Solitons and Fractals} {\bf28} 100--111.

%\bibitem{kn:MAM08} Mart{\'i}nez, G. J., Adamatzky, A., \& McIntosh, H. V. (2008) On the representation of gliders in Rule 54 by de Bruijn and cycle diagrams, \textit{Lecture Notes in Computer Science} {\bf 5191} 83--91.

%\bibitem{kn:MAM10} Mart{\'i}nez, G. J., Adamatzky, A., \& McIntosh, H. V. (2010) Localization dynamics in a binary two-dimensional cellular automaton: the Diffusion Rule, \textit{Journal of Cellular Automata} {\bf 5(4-5)} 289--313.

%\bibitem{kn:MAMM10} Mart{\'i}nez, G. J., Adamatzky, A., Morita, K., \& Margenstern, M. (2010) Computation with competing patterns in Life-like automaton, In: \textit{Game of Life Cellular Automata}, A. Adamatzky (Ed.), Springer, 547--572.

\bibitem{kn:Mar07} Margenstern, M. (2007) \textit{Cellular Automata in Hyperbolic Spaces}, Old City Publishing, Inc.

\bibitem{kn:Marg84} Margolus, N. H. (1984) Physics-like models of computation, \textit{Physica D} {\bf 10 (1-2)} 81--95.

\bibitem{kn:Marg98} Margolus, N. H. (1998) Crystalline Computation, In: \textit{Feynman and computation: exploring the limits of computers}, A. J. G. Hey (Ed.), Perseus Books, 267--305.

\bibitem{kn:Marg03} Margolus, N. H. (2003) Universal Cellular Automata Based on the Collisions of Soft Spheres, In: \textit{New constructions in cellular automata}, D. Griffeath \& C. Moore (Eds.), Oxford University Press, 231--260.

%\bibitem{kn:Mar00} Martin, B. (2000) A Group Interpretation of Particles Generated by One-Dimensional Cellular Automaton, Wolfram's Rule 54, \textit{Int. J. Modern Physics C} {\bf 11(1)} 101--123.

\bibitem{kn:MAS10} Mart{\'i}nez, G. J., Adamatzky, A., Seck-Tuoh-Mora, J. C., \& Alonso-Sanz, A. (2010) How to make dull cellular automata complex by adding memory: Rule 126 case study, \textit{Complexity} {\bf 15(6)} 34--49.

\bibitem{kn:MMAM10} Mart{\'i}nez, G. J., Morita, K., Adamatzky, A. \& Margenstern, M. (2010) Majority adder implementation by competing patterns in Life-like rule $B2/S2345$, \textit{Lecture Notes in Computer Science} {\bf 6079} 93--104.

\bibitem{kn:MMS06} Mart{\'i}nez, G. J., McIntosh, H. V., \& Seck-Tuoh-Mora, J. C. (2006) Gliders in Rule 110, \textit{Int. J. of Unconventional Computing} {\bf 2(1)} 1--49.

%\bibitem{kn:MMS07} Mart{\'i}nez, G. J., McIntosh, H. V., Seck-Tuoh-Mora, J. C., \& Chapa-Vergara, S. V. (2007) Rule 110 objects and other constructions based-collisions, \textit{Journal of Cellular Automata} {\bf 2(3)} 219--242.

%\bibitem{kn:MMS08} Mart{\'i}nez, G. J., McIntosh, H. V., Seck-Tuoh-Mora, J. C., \& Chapa-Vergara, S. V. (2008) Determining a regular language by glider-based structures called {\it phases f$_i$\_1} in Rule 110, \textit{Journal of Cellular Automata} {\bf 3(3)} 231--270.

\bibitem{kn:MMS11} Mart{\'i}nez, G. J., McIntosh, H. V., Seck-Tuoh-Mora, J. C., \& Chapa-Vergara, S. V. (2011) Reproducing the cyclic tag system developed by Matthew Cook with Rule 110 using the phases f$_1$\_1,'' \textit{Journal of Cellular Automata} {\bf 6(2-3)} 121--161.

\bibitem{kn:MAS11} Mart{\'i}nez, G. J., Adamatzky, A., Stephens, C. R., \& Hoeflich, A. F. (2011) Cellular automaton supercolliders, \textit{International Journal of Modern Physics C} {\bf 22(4)} 419--439.

\bibitem{kn:Mc90} McIntosh, H. V. (1990) Wolfram's Class IV and a Good Life, \textit{Physica D} {\bf 45} 105--121.

\bibitem{kn:Mc91} McIntosh, H. V. (1991) Linear Cellular Automata via de Bruijn Diagrams, \url{http://delta.cs.cinvestav.mx/~mcintosh/oldweb/cf/debruijn.html}.

\bibitem{kn:Mc91a} McIntosh, H. V. (1991) Reversible Cellular Automata, \url{http://delta.cs.cinvestav.mx/~mcintosh/oldweb/ra/ra.html}.

\bibitem{kn:Mc99} McIntosh, H. V. (1999) Rule 110 as it relates to the presence of gliders, \url{http://delta.cs.cinvestav.mx/~mcintosh/comun/RULE110W/RULE110.html}.

\bibitem{kn:Mc00} McIntosh, H. V. (2000) A Concordance for Rule 110, \url{http://delta.cs.cinvestav.mx/~mcintosh/comun/ccord/ccord.html}.

\bibitem{kn:Mc02} McIntosh, H. V. (2002) Rule 110 Is Universal!, \url{http://delta.cs.cinvestav.mx/~mcintosh/comun/texlet/texlet.html}.

\bibitem{kn:Mc09} McIntosh, H. V. (2009) \textit{One Dimensional Cellular Automata}, Luniver Press.

\bibitem{kn:MH89} Morita, K. \& Harao, M. (1989) Computation universality of one-dimensional reversible (injective) cellular automata, \textit{Trans. IEICE Japan} {\bf E-72} 758--762. 

\bibitem{kn:MHC93} Mitchell, M., Hraber, P. T., \& Crutchfield, J. P. (1993) Revisiting the Edge of Chaos: Evolving Cellular Automata to Perform Computations, \textit{Complex Systems} {\bf 7(2)} 89--130.

\bibitem{kn:Mills08} Mills, J. W. (2008) The Nature of the Extended Analog Computer, \textit{Physica D} {\bf 237(9)} 1235--1256.

\bibitem{kn:MMI99} Morita, K., Margenstern, M., \& Imai, K. (1999) Universality of reversible hexagonal cellular automata, \textit{Theoret. Informatics Appl.} {\bf 33} 535--550.


\bibitem{kn:Mor90} Morita, K. (1990) A simple construction method of a reversible finite automaton out of Fredkin gates, and its related problem, \textit{Trans. IEICE Japan} {\bf E-73} 978--984.

\bibitem{kn:Mor07} Morita, K. (2007) Simple universal one-dimensional reversible cellular automata, \textit{Journal of Cellular Automata} {\bf 2} 159--166.

\bibitem{kn:Mor08} Morita, K. (2008) Reversible computing and cellular automata---A survey, \textit{Theoretical Computer Science} {\bf 395} 101--131

\bibitem{kn:MTI02} Morita, K., Tojima, Y., Imai, K. \& Ogiro, T. (2002) Universal Computing in Reversible and Number-Conserving Two-Dimensional Cellular Spaces, In: \textit{Collision-Based Computing}, A. Adamatzky (Ed.), Springer, 161--199.

\bibitem{kn:MTV86} Margolus, N., Toffoli, T., \& Vichniac, G. (1986) Cellular-Automata Supercomputers for Fluid Dynamics Modeling, \textit{Physical Review Letters} {\bf 56(16)} 1694--1696.

\bibitem{kn:Nasu78} Nasu, M. (1978) Local Maps Inducing Surjective Global Maps of One-Dimensional Tesselation Automata, \textit{Mathematical Systems Theory} {\bf 11} 327--351

\bibitem{kn:PST86} Park, J. K., Steiglitz, K. \& Thurston, W. P. (1986) Soliton-like behaviour in automata, \textit{Physica D} {\bf 19} 423--432.

\bibitem{kn:Ren02} Rendell, P. (2002) Turing universality of the game of life, In: \textit{Collision-Based Computing}, A. Adamatzky (Ed.), Springer, 513--540.

\bibitem{kn:Ren11} Rendell, P. (2011) A Universal Turing Machine in Conway's Game of Life, \textit{Proceedings of the 2011 International Conference on High Performance Computing \& Simulation}, IEEE Xplore 10.1109/HPCSim.2011.5999906, 764--772.

%\bibitem{kn:Renn02} Rennard, J. P. (2002) Implementation of Logical Functions in the Game of Life, In: \textit{Collision-Based Computing}, A. Adamatzky (Ed.), Springer, 491--512.

\bibitem{kn:SBC04} Sapin, E., Bailleux, O., Chabrier, J.-J. \& Collet, P. (2004) A New Universal Cellular Automaton Discovered by Evolutionary Algorithms, \textit{Lecture Notes in Computer Sciences} {\bf 3102} 175--187.

\bibitem{kn:SBC07} Sapin, E., Bailleux, O., Chabrier, J.-J. \& Collet, P. (2007) Demonstration of the Universality of a New Universal Cellular Automaton, \textit{Int. J. Unconventional Computing} {\bf 3(2)} 79--103.

\bibitem{kn:SCM05} Seck-Tuoh-Mora, J. C., Chapa-Vergara, S. V., Mart{\'i}nez, G. J., \& McIntosh, H. V. (2005) Procedures for calculating reversible one-dimensional cellular automata, \textit{Physica D} {\bf 202} 134--141.

\bibitem{kn:SHM08} Seck-Tuoh-Mora, J. C., Hern\'andez, M. G., McIntosh, H. V., \& Chapa-Vergara, S. V. (2008) The Inverse behaviour of a Reversible One-Dimensional Cellular Automaton Obtained by a Single Welch Diagram, In: \textit{Automata 2008: Theory and Applications of Cellular Automata}, A. Adamatzky, R. A. Sanz, A. Lawniczak, G. J. Mart{\'i}nez, K. Morita, \& T. Worsh (Eds.), Luniver Press, 114--125.

\bibitem{kn:SMM06} Seck-Tuoh-Mora, J. C., Mart{\'i}nez, G. J., \& McIntosh, H. V. (2006) The Inverse behaviour of a Reversible One-Dimensional Cellular Automaton Obtained by a Single Welch Diagram, \textit{Journal of Cellular Automata} {\bf 1(1)} 25--39.

\bibitem{kn:Sipp97} Sipper, M. (1997) \textit{Evolution of Parallel Cellular Machines: The Cellular Programming Approach}, Springer.

\bibitem{kn:Siw02} Siwak, P. (2002) Iterons of Automata, In: \textit{Collision-Based Computing}, A. Adamatzky (Ed.), Springer, 299--354.

\bibitem{kn:Sut89} Sutner, K. (1989) A note on Culik-Yu classes, \textit{Complex Systems} {\bf 3(1)} 107--115.

\bibitem{kn:Sut91} Sutner, K. (1991) De Bruijn graphs and linear cellular automata, \textit{Complex Systems} {\bf 5(1)} 19--30.

\bibitem{kn:Sut09} Sutner, K. (2009) Classification of Cellular Automata, In \textit{Encyclopedia of Complexity and Systems Science}, Part 3, Springer, R. A Meyers (Ed.), 755--768.

\bibitem{kn:Suzu94} Suzudo, T. (1994) Spatial pattern formation in asynchronous cellular automata with mass conservation, \textit{Physica A} {\bf 343} 185--200.

\bibitem{kn:Toff98} Toffoli, T. (1998) Non-Conventional Computers, In: \textit{Encyclopedia of Electrical and Electronics Engineering}, J. Webster (Ed.), Wiley \& Sons, 455--471.

\bibitem{kn:TM87} Tommaso, T. \& Norman, M. (1987) \textit{Cellular Automata Machines}, The MIT Press, Cambridge, Massachusetts.

\bibitem{kn:TV02} Tomassini, M. \& Venzi, M. (2002) Evolving Robust Asynchronous Cellular Automata for the Density Task \textit{Complex Systems} {\bf 13(1)} 185--204.

\bibitem{israeli} Israeli, N. and Goldenfeld, N. Computational Irreducibility and the Predictability of Complex Physical Systems, \textit{Phys. Rev. Lett.} 92, 74105--74108, 2004.

\bibitem{kn:Voor96} Voorhees, B.H. (1996) \textit{Computational analysis of one-dimensional cellular automata}, World Scientific Series on Nonlinear Science, Series A, Vol. 15.

\bibitem{kn:Voor06} Voorhees, B.H. (2008) Remarks on Applications of  De Bruijn Diagrams and Their Fragments, \textit{Journal of Cellular Automata} {\bf 3(3)} 187--204.

\bibitem{kn:von66} von Neumann, J. (1966) \textit{Theory of Self-reproducing Automata} (edited and completed by A. W. Burks), University of Illinois Press, Urbana and London.

\bibitem{kn:WL92} Wuensche, A. \& Lesser, M. (1992) \textit{The Global Dynamics of Cellular Automata}, Addison-Wesley Publishing Company.

\bibitem{kn:WN09} Worsch, T. \& Nishio, H. (2009) Achieving Universality of CA by Changing the neighbourhood, \textit{Journal of Cellular Automata} {\bf 4(3)} 237--246.

\bibitem{kn:Wolf83} Wolfram, S. (1983) Statistical Mechanics of Cellular automata, \textit{Reviews of Modern Physics} {\bf 55(3)} 601--644.

\bibitem{kn:Wolf84} Wolfram, S. (1984) Universality and complexity in cellular automata, \textit{Physica D} {\bf 10} 1--35.

\bibitem{kn:Wolf84a} Wolfram, S. (1984) Computation Theory of Cellular Automata, \textit{Communications in Mathematical Physics} {\bf 96} 15--57.

\bibitem{kn:Wolf86} Wolfram, S. (1986) Random Sequence Generation by Cellular Automata, \textit{Advances in Applied Mathematics} {\bf 7} 123--169.

\bibitem{kn:Wolf88} Wolfram, S. (1988) Cellular Automata Supercomputing, In: \textit{High Speed Computing: Scientific Applications and Algorithm Design}, R. B. Wilhelmson (Ed.), University of Illinois Press, 40--48.

\bibitem{kn:Wolf94} Wolfram, S. (1994) \textit{Cellular Automata and Complexity}, Addison-Wesley Publishing Company.

\bibitem{wolframopen} Wolfram, S. (1985) Twenty Problems in the Theory of Cellular Automata, \textit{Physica Scripta}, T9, 170--183.

\bibitem{kn:Wolf02} Wolfram, S. (2002) \textit{A New Kind of Science}, Wolfram Media, Inc., Champaign, Illinois.

\bibitem{kn:Wors09} Worsch, T. (2009) Cellular Automata as Models of Parallel Computation, In \textit{Encyclopedia of Complexity and Systems Science}, R. A. Meyers (Ed.), Springer, 741--755.

\bibitem{kn:WO08} Wolz, D. \& de Oliveira, P. P. B. (2008) Very effective evolutionary techniques for searching cellular automata rule
spaces, \textit{Journal of Cellular Automata} {\bf 3(4)} 289--312.

\bibitem{kn:Wue98} Wuensche, A. (1998) Genomic Regulation Modeled as a Network with Basins of Attraction, In: \textit{Pacific Symposium on Biocomputing '98}, R. B. Altman, A. K. Dunker, L. Hunter, \& T. E. Klien (Eds.) World Scientific, Singapore, 89--102.

\bibitem{kn:Wue99} Wuensche, A. (1999) Classifying Cellular Automata Automatically, \textit{Complexity} {\bf 4(3)} 47--66.

\bibitem{kn:Zen10} Zenil H. (2010), Compression-based Investigation of the Dynamical Properties of Cellular Automata and Other Systems, \textit{Complex Systems,} vol. 19, No. 1, 1--28.

\bibitem{kn:ZenAISB} Zenil, H. (2012), Nature-like Computation and a Measure of Computability, In Dodig-Crnkovic, G. and Giovagnoli, R. (eds), \textit{Natural Computing/Unconventional Computing and its Philosophical Significance}, SAPERE Series, Springer.

\bibitem{kn:Zen12} Zenil H. (2012), On the Dynamic Qualitative behaviour of Universal Computation, \textit{Complex Systems,} vol. 20, No. 3, 265--278.

\bibitem{kn:ZenJETAI} Zenil H. (forthcoming), Programmability Tests for Natural Computation with the Game of Life as a Case Study, \textit{Journal of Experimental and Theoretical Artificial Intelligence}, (accepted).

\end{thebibliography}
\end{document}